\documentclass[aps, prb, twocolumn, groupedaddress, showpacs, floatfix]{revtex4-1}
\usepackage{amsmath}
\usepackage{amssymb}
\usepackage{graphicx}
\usepackage{xr-hyper}
\usepackage{bm}

\begin{document}
\title{Inelastic Current Noise in Nanoscale Systems: Scattering Theory Analysis}

\author{Sejoong Kim}
\email[ ]{sejoong@kias.re.kr}
\affiliation{Korea Institute for Advanced Study, Seoul 130-722, Korea}

\date{\today}

\begin{abstract}
We present a scattering theory description for the inelastic current noise in the presence of electron-vibration interactions. In this description, we specify elastic and inelastic scattering contributions to the shot noise by examining charge transfers between scattering states and energy exchange between electrons and vibrations. The elastic and inelastic scattering processes are further decomposed into current correlations of electrons at the same energy and those of electrons at different energies. Focusing on the inelastic noise signals defined as steps in the voltage derivative of the shot noise, we show that single-channel systems have two ranges of transmission at which the inelastic noise signals exhibit the crossover between positive and negative signs. In a high transmission regime, even and odd vibrational modes of mirror-symmetric systems provide upper and lower bounds to the ratio of the inelastic noise signal to the conductance step. This can be a theoretical justification for models used to understand the recent noise experiment [Phys. Rev. Lett. \textbf{108}, 146602 (2012)] and numerical calculations on gold atomic chains [Phys. Rev. B \textbf{86}, 155411 (2012)]. 
\end{abstract}

\pacs{72.10.-d, 72.10.Di, 72.70.+m, 73.63.-b}

\maketitle


%
%
\section{Introduction}
Since pioneering experiments measuring inelastic electron tunneling spectroscopy (IETS) signals~\cite{PRL1966Lambe, Science1998Stipe, PRL2000Hahn, Nature2002Smit, PRL2002Agrait}, intense efforts have been made to understand inelastic transport properties when conducting electrons interact with local vibrations~\cite{NanoLett2004Wang, PRB2005Djukic, PRL2006Thijssen, PRL2008Tal, PRB1970Davis, JPhysC1972Caroli, PRL1987Persson, PRL1995Bonca, PRL1999Ness, PRB2000Emberly, PRL2000Lorente, PRL2001Lorente, 2003Montgomery1, 2003Montgomery2, NanoLett2003Chen, NanoLett2004Chen, NanoLett2005Chen, NanoLett2005Jiang, JChP2004Galperin, SurfSci2007Ueba, PRB2005Viljas, PRL2004Frederiksen, PRB2005Paulsson, PRB2006Vega, PRL2008Paulsson, PRB2007Frederiksen_a, PRB2007Frederiksen_b, PRB2009Kristensen, PRB2013Kim, PRB2013SKim, 2008PRBEgger, 2009PRBEntin-Wohlman, PRB2007Gagliardi, JChemPhys2006Troisi, PRL2005Chen, PRB2009Avriller, PRB2009Schmidt, PRL2009Haupt, PRB2010Haupt, PRB2010Urban, PRB2011Novotny, PRB2011Park, PRB2012Avriller, PRL2012Kumar, NatComm2010Tsutsui}. The IETS signals are identified as steps in the differential conductance at a threshold bias voltage equal to a vibrational energy~\cite{PRB2005Viljas, PRB2006Vega, PRL2008Paulsson, PRL2004Frederiksen, PRB2005Paulsson, PRB2007Frederiksen_a, PRB2007Frederiksen_b, PRB2009Kristensen, PRB2013Kim, PRB2013SKim, PRB2007Gagliardi,JChemPhys2006Troisi}. The same steps also appear in the voltage derivative of the shot noise~\cite{PRL2005Chen, PRB2009Avriller, PRB2009Schmidt, PRL2009Haupt, PRB2010Haupt, PRB2010Urban, PRB2011Novotny, PRB2011Park, PRB2012Avriller, PRL2012Kumar, NatComm2010Tsutsui}. These steps indicate opening of inelastic transport channels, where electrons can pass through the junction by losing energy.

It is known that the conductance variation undergoes a crossover from an increase to a decrease when a bare transmission $\mathcal{T}$ evolves from zero to one~\cite{PRB2006Vega, PRB2005Paulsson, PRL2008Paulsson}. The similar crossover behavior in the shot noise has been reported in a recent experiment of the shot noise measurement~\cite{PRL2012Kumar}. In Ref.~\onlinecite{PRL2012Kumar}, it is observed that inelastic noise corrections are negative when a measured zero-bias conductance is approximately below $0.95 G_{0}$. In contrast, the noise corrections are exclusively positive for the zero-bias conductance close to $G_{0}$. 

To understand this crossover, several model systems have been theoretically investigated~\cite{PRL2012Kumar,PRB2012Avriller}. 
The representative system is a single-level model with a single vibrating scatterer~\cite{PRB2009Avriller, PRB2009Schmidt, PRL2009Haupt, PRL2012Kumar}. It is shown that the conductance crossover occurs when $\mathcal{T}=0.5$ for the single level symmetrically connected to single-channel electrodes~\cite{PRL2008Paulsson}. Considering the current noise, the single-level model symmetrically coupled to electrodes exhibits two crossovers in the inelastic noise correction at $\mathcal{T}=(2\pm\sqrt{2})/4$~\cite{PRB2009Avriller, PRB2009Schmidt, PRL2009Haupt, PRL2012Kumar}. Reference~\onlinecite{PRB2012Avriller} studied a two-site tight-binding model symmetrically connected to electrodes in order to analyze density functional theory (DFT) calculations on inelastic signals of Au atomic point contacts. For the out-of-phase longitudinal vibrational mode that gives a dominant contribution to inelastic signals, it is found that the two-site model exhibits the crossover in the conductance at $\mathcal{T}=0.5$ and the noise signal crossover at $\mathcal{T}=(2\pm\sqrt{2})/4$, as the symmetric single-level model does~\cite{PRB2012Avriller}. 
Although these simple models can give some qualitative hint on understanding the inelastic shot noise, it has not yet been clearly understood why the models exhibit the same crossover behaviors. Furthermore, performing DFT calculations for a variety of Au atomic chains, Ref.~\onlinecite{PRB2012Avriller} calculated the ratios of the steps in the inelastic noise signal to the conductance changes for the chains, and compared them with the ratio of the two-site model. Reference~\onlinecite{PRB2012Avriller} concluded that the ratio predicted by the two-site model might be an upper bound to those of the atomic chains. However, there is no theoretical justification for why the two-site model, which is originally designed to study atomic point contacts, can provide an upper bound to the computed ratios of the atomic chains varying in configurations and lengths. Thus, it is needed to find out a general picture that can explain the aforementioned issues. 


Recently we presented a scattering-state description of inelastic electron transport in a weak electron-vibration (el-vib) coupling regime, which is established by converting non-equilibrium Green's functions (NEGF) to scattering states~\cite{PRB2013SKim}. In this description, we clarified elastic and inelastic contributions to conductance variations in terms of eigenchannel scattering states and associated scattering matrices. Doing so, we obtained a general expression for the conductance jump of single-channel systems. While the inelastic contribution is proportional to $1-2\mathcal{T}$, the elastic one is always negative. It leads to conclusion that the conductance crossover can generally occur below $\mathcal{T}=0.5$ for single-channel systems.  

In this paper, we extend our scattering theory approach to the inelastic current noise, especially in order to unveil an unified picture that can comprehensively explain the existing results in the literature~\cite{PRL2012Kumar, PRB2012Avriller}. We specify elastic and inelastic contributions to the current noise by examining charge transfers between scattering states and energy exchange between electrons and vibrations. Different from the conductance, such contributions to the current noise can be further decomposed into current correlations of electrons at the same energy and those of electrons at different energies. Considering single-channel systems, we show that the crossover for the inelastic noise signal can occur in two ranges of transmissions, {\it i.e.}, between $0$ and $(2-\sqrt{2})/4$ or between $0.5$ and $(2+\sqrt{2})/4$. In particular, for mirror-symmetric junctions, even parity vibrational modes lead to two crossovers of the noise correction at $\mathcal{T}=(2\pm\sqrt{2})/4$, while there is only one crossover at $\mathcal{T}=0.5$ for odd parity modes. 
When $\mathcal{T}$ is close to one, our scattering-state description shows that 
the ratios of noise correction steps to conductance changes for odd and even parity modes are lower and upper bounds for those of general cases, respectively.

This paper is organized as follows. In Sec.~\ref{general_picture}, we present a general argument about how elastic and inelastic scattering processes compete with each other for transport properties such as conductance and shot noise. In Sec.~\ref{NEGF_method}, we summarize current and shot noise expressions in the NEGF formalism~\cite{PRB2010Haupt}. In Sec.~\ref{scattering_theory}, we present our scattering-theory description of inelastic transport. First, we briefly summarize the current correction due to el-vib scatterings discussed in our previous work~\cite{PRB2013SKim}. Second, we explain the scattering-state description of the inelastic noise correction, focusing on identification of scattering processes in the noise correction. In Sec.~\ref{discussion}, we apply our theory to single-channel systems. Here we discuss the crossover of the noise correction in general situations including the systems investigated in the literature~\cite{PRB2012Avriller}. In Sec.~\ref{conclusion}, we make a final conclusion. Technical details and derivations are discussed in Appendixes~\ref{derivation} and \ref{multichannel_systems}.

%
%
\section{Competition between Elastic and Inelastic processes}\label{general_picture}
Before presenting our scattering-theory description for the shot noise, we first discuss the competition between elastic and inelastic scattering processes in a general perspective~\cite{PRB1970Davis, JPhysC1972Caroli, JChP2004Galperin, SurfSci2007Ueba, PRB2005Viljas, PRB2006Vega, PRL2008Paulsson, PRB2009Avriller}. Here we define that electron transport is elastic (inelastic) if the total energy of conducting electrons is conserved (not conserved) after electrons pass through the vibrating region. In Ref.~\onlinecite{PRB2013SKim}, we clarified elastic and inelastic scattering contributions to the current in the second order of el-vib couplings, and used their interplay to explain the conductance crossover. 
In fact, the competition between elastic and inelastic scattering processes is not limited only to the current correction~\cite{PRB1970Davis, JPhysC1972Caroli, JChP2004Galperin, SurfSci2007Ueba, PRB2005Viljas, PRB2006Vega, PRL2008Paulsson}, but it is generally respected in other transport properties such as the current noise~\cite{PRB2009Avriller}. 
When a zero temperature regime is considered for simplicity, a many-body state of electrons and vibrons perturbed by el-vib interactions can be written as
\begin{equation}
|\Psi\rangle = |\Psi_{0}\rangle + |\Psi_{1}\rangle + |\Psi_{2}\rangle \cdots,
\end{equation}
where $|\Psi_{0}\rangle$ is a ground state of no el-vib coupling at zero temperature, {\it i.e.}, $|\Psi_{0}\rangle=|\psi_{0}\rangle_{e} \otimes |0\rangle_{v}$ where $|\psi_{0}\rangle_{e}$ and $|0\rangle_{v}$ denote electronic and vibronic ground states respectively. Note that the vibronic state $|n\rangle_{v}$ represents that there are $n$ vibrons excited. 
$|\Psi_{n}\rangle\;(n=1,2, \cdots)$ is a perturbed state in the $n$th order of el-vib couplings. 
Since electrons emit one vibron when electrons interact with vibrations once, the first-order perturbed state is written as $|\Psi_{1}\rangle=|\psi_{1}\rangle_{e} \otimes |1\rangle_{v}$, where $|\psi_{1}\rangle_{e}$ denotes the electronic state after losing its energy by one vibronic quantum $\hbar\omega_{\lambda}$.
At the second order of the el-vib coupling, two cases are possible: The first case is that electrons regain the vibronic energy quantum from vibrations, and the other is that electrons emit another vibron. 
Thus, the second-order perturbed state is expressed as a superposition of these two possibilities, {\it i.e.}, $|\Psi_{2}\rangle = |\psi_{2}^{(1)}\rangle_{e} \otimes |0\rangle_{v} + |\psi_{2}^{(2)}\rangle_{e} \otimes |2\rangle_{v}$.
Here $|\psi_{2}^{(1)}\rangle_{e}$ and $|\psi_{2}^{(2)}\rangle_{e}$ denote electronic wavefunctions associated with the first case (emission-reabsorption) and the second one (double-vibron emission) respectively.
For any electronic operator $\mathcal{O}$ that does not change the vibronic occupation, it is shown that the expectation value of $\mathcal{O}$ with respect to $|\Psi\rangle$ is expanded as
\begin{equation}
\langle \mathcal{O} \rangle = \frac{\langle \Psi |\mathcal{O}| \Psi \rangle }{\langle \Psi|\Psi \rangle} = \langle\mathcal{O}\rangle_{0} + \langle\mathcal{O}\rangle_{2}^{\textrm{in}} + \langle\mathcal{O}\rangle_{2}^{\textrm{el}} + \cdots,
\end{equation}
where $\langle\mathcal{O}\rangle_{0} = \langle\Psi_{0}|\mathcal{O}|\Psi_{0}\rangle$, and 
\begin{eqnarray}
\label{inelastic_general}\langle\mathcal{O}\rangle_{2}^{\textrm{in}} &=& \langle\Psi_{1}|\mathcal{O}|\Psi_{1}\rangle -\langle\Psi_{1}|\Psi_{1}\rangle\langle\Psi_{0}|\mathcal{O}|\Psi_{0}\rangle, \\
\label{elastic_general}\langle\mathcal{O}\rangle_{2}^{\textrm{el}} &=& 2\textrm{Re}\left[\langle\Psi_{0}|\mathcal{O}|\Psi_{2}\rangle\right].
\end{eqnarray}
Here Eq.~(\ref{inelastic_general}) is the inelastic correction since it includes only $|\Psi_{1}\rangle$ where electrons lose energy by emitting one vibron.
Since the vibronic state $|2\rangle_{v}$ associated with $|\psi^{(2)}_{2}\rangle_{e}$ is orthogonal to $|0\rangle_{v}$ of the ground state, $|\psi^{(2)}_{2}\rangle_{e}$ does not contribute to $\langle \mathcal{O}\rangle^{\textrm{el}}_{2}$. Thus, Eq.~(\ref{elastic_general}) is reduced to $2\textrm{Re}[ _{e}\langle\psi_{0}|\mathcal{O}|\psi^{(1)}_{2}\rangle_{e} ]$, 
thereby implying that it is the elastic process. 
Since the elastic correction Eq.~(\ref{elastic_general}) is a form of interference between $|\psi_{0}\rangle_{e}$ and $|\psi_{2}^{(1)}\rangle_{e}$, it is expected to contain some phase information. In fact, the elastic corrections to the current and the noise depend on phases of scattering matrix elements and transition amplitude between scattering states as shown later. Based on this general structure that elastic and inelastic scattering processes compete with each other in the nontrivial lowest order of el-vib couplings, we will discuss the inelastic current noise in detail in the following sections.

%
%
\section{Nonequilibrium Green's Function Method}\label{NEGF_method}
The system in which we are interested is the two-terminal geometry with local vibrations located in a conductor region. 
The corresponding Hamiltonian is written as 
\begin{equation}
\mathcal{H} = \mathcal{H}_{el} + \mathcal{H}_{vib} + \mathcal{H}_{el-vib},
\end{equation}
where $\mathcal{H}_{el}$ and $\mathcal{H}_{vib}$ are the electronic Hamiltonian describing the two-terminal setup and the vibronic one of local vibrations respectively. 
$\mathcal{H}_{el-vib}$ is the coupling term between electronic and vibronic Hamiltonians.  The electronic Hamiltonian $\mathcal{H}_{el}$ is further divided into
two electrodes (left and right) $H_{L/R}$, a central part $H_{C}$ connected to leads, and a coupling $H_{T}$ between leads and the central device. 
In the second quantized form, it is 
\begin{equation}
\mathcal{H}_{el} = H_{L} + H_{R} + H_{C}  + H_{T},
\end{equation}
where 
\begin{subequations}
\begin{eqnarray}
H_{\alpha = L,R} &=& \sum_{k} \epsilon_{\alpha k} c^{\dagger}_{\alpha k} c_{\alpha k} \\
H_{C}            &=& \sum_{i,j} \varepsilon_{ij} d^{\dagger}_{i} d_{j} \\
H_{T}            &=& \sum_{\alpha=L,R}\sum_{k,i} V_{\alpha k i} c^{\dagger}_{\alpha k} d_{i} + \textrm{H.c.}
\end{eqnarray} 
\end{subequations}
Here $c^{\dagger}_{\alpha k}$ ($c_{\alpha k}$) and $d^{\dagger}_{i}$ ($d_{i}$) represent electronic creation (annihilation) operators for the electrodes and the device part, respectively. Note that a spin index is not explicitly shown. 
$\mathcal{H}_{vib}$ describes local vibrations confined in the device region, which consists of a collection of harmonic oscillators,
\begin{equation}
\mathcal{H}_{vib} = \sum_{\lambda}\hbar\omega_{\lambda} a_{\lambda}^{\dagger} a_{\lambda},
\end{equation}
where $a^{\dagger}_{\lambda}$ ($a_{\lambda}$) is a vibronic creation (annihilation) operator for the $\lambda$th mode. 
The coupling Hamiltonian between electrons and local vibrations is given by 
\begin{equation}
\mathcal{H}_{el-vib} = \sum_{\lambda} \sum_{i,j} \mathcal{M}^{\lambda}_{ij} d^{\dagger}_{i} d_{j} \left(a_{\lambda} + a_{\lambda}^{\dagger} \right).
\end{equation}

Next we consider the non-equilibrium transport theory based on NEGFs in a weak el-vib coupling regime~\cite{JPhysC1972Caroli, JChP2004Galperin, SurfSci2007Ueba, PRB2005Viljas, PRB2006Vega, PRL2008Paulsson, PRB2005Paulsson, PRB2007Frederiksen_a, PRL2004Frederiksen, PRB2007Frederiksen_b, PRB2009Kristensen}. 
In a zero temperature limit and a regime where a damping rate of vibrations is much larger than a heating rate, 
the current correction due to el-vib interactions is 
\begin{subequations}
\begin{eqnarray}
\delta I &=&  \frac{2e}{h} \sum_{\lambda} \int^{\mu_{L}}_{\mu_{R}} d\varepsilon 2\textrm{ReTr}\left[C \mathcal{M}^{\lambda} \textrm{Re}G^{r,-}_{0} \mathcal{M}^{\lambda} \right] \label{qel_current} \\
         & & +\frac{2e}{h} \sum_{\lambda} \int_{\mu_{R}+\hbar\omega_{\lambda}}^{\mu_{L}}d\varepsilon\textrm{Tr}\left[B\mathcal{M}^{\lambda}A_{R}^{-} \mathcal{M}^{\lambda}\right] \\
         & & +\frac{2e}{h} \sum_{\lambda} \int^{\mu_{L}}_{\mu_{R}+\hbar\omega_{\lambda}} d\varepsilon \textrm{Im} \textrm{Tr} \left[ C \mathcal{M}^{\lambda} A_{R}^{-} \mathcal{M}^{\lambda} \right] \\
         & & +\frac{2e}{h} \sum_{\lambda} \int^{\mu_{L}-\hbar\omega_{\lambda}}_{\mu_{R}} d\varepsilon \textrm{Im} \textrm{Tr} \left[ C \mathcal{M}^{\lambda} A_{L}^{+} \mathcal{M}^{\lambda} \right],
\end{eqnarray} 
\end{subequations}
where $B \equiv G^a_{0} \Gamma_{L} G^r_{0}$, $C \equiv A_{R} \Gamma_{L} G^r_{0}$, and $D \equiv A_{R} \Gamma_{L} A_{R}$. 
$G^{r(a)}_{0}$ is the retarded (advanced) Green's function~\cite{1995Datta} of the conductor part without el-vib interactions, 
which is expressed as
$G^{r(a)}_{0}(\varepsilon) = [\varepsilon - H_{C} - \Sigma^{r(a)}_{L} -\Sigma^{r(a)}_{R}]^{-1}$.
Here $\Sigma^{r(a)}_{\alpha}$ denotes the retarded (advanced) lead self-energy of the electrode $\alpha=L,R$.
$\Gamma_{\alpha}=i[\Sigma^{r}_{\alpha}-\Sigma^{a}_{\alpha}]$ is the coupling function to leads,
and $A_{\alpha} = G_{0}^{r} \Gamma_{\alpha} G^{a}_{0}$ is the spectral function of the conductor region originating from the electrode $\alpha=L,R$~\cite{1995Datta}. 
Superscripts $\pm$ indicate that energy argument is $\varepsilon \pm \hbar \omega_{\lambda}$ and without $\pm$, the argument is $\varepsilon$. 
We assume that the left chemical potential $\mu_{L}$ is bigger than the right one $\mu_{R}$.  
The first term [Eq.~(\ref{qel_current})] is interpreted as a quasi-elastic correction to a bare transmission~\cite{PRB2010Haupt}. 
Since Eq.~(\ref{qel_current}) does not contribute to the conductance step at the threshold bias voltage in the regime of our interest, it is neglected in the following discussion.  

To the second order of el-vib interactions, the shot noise correction $\delta S$ has been derived by using the counting field method in the full-counting statistics~\cite{PRL2009Haupt, PRB2009Avriller, PRB2009Schmidt, PRB2010Haupt}. 
The inelastic noise correction in the single-level model was first investigated by several groups~\cite{PRB2009Avriller, PRB2009Schmidt, PRL2009Haupt}, and later it was extended to general cases where multiple electronic levels and many vibrational modes are involved in a regime of equilibriated vibrations~\cite{PRB2010Haupt}. 
Recently the backaction effect of nonequilibrium vibrational populations on the shot noise has been investigated~\cite{PRB2010Urban, PRB2011Novotny}. Following Ref.~\onlinecite{PRB2010Haupt}, we consider situations where multiple electronic levels are coupled to many vibrational modes in the regime of zero temperature and equilibriated vibrons.
The noise correction $\delta S$ has two contributions $\delta S_{\textrm{mf}}$ and $\delta S_{\textrm{vc}}$, which are classified as {\it mean-field} and {\it vertex} corrections in the diagrammatic representation~\cite{PRB2010Haupt}: 
\begin{subequations}
\begin{eqnarray}
\label{qel_noise} \frac{\delta S_{\textrm{mf}}}{2e^2/h} &=&  \sum_{\lambda}\int^{\mu_{L}}_{\mu_{R}} 2\textrm{ReTr}\left[\left(1-2T\right)C\mathcal{M}^{\lambda}\textrm{Re}G^{r,-}_{0}\mathcal{M}^{\lambda}\right] \\
\label{noise_mf1}                                       & & +\sum_{\lambda}\int_{\mu_{R}+\hbar\omega}^{\mu_{L}}d\varepsilon\textrm{Tr}\left[\left(1-2T\right)G^{r}_{0}\mathcal{M}^{\lambda}A_{R}^{-}\mathcal{M}^{\lambda}G^{a}_{0}\Gamma_{L}\right] \nonumber \\ \\
\label{noise_mf2}                                       & & +\sum_{\lambda}\int_{\mu_{R}+\hbar\omega}^{\mu_{L}}d\varepsilon\textrm{Im}\textrm{Tr}\left[\left(1-2T\right)C\mathcal{M}^{\lambda}A_{R}^{-}\mathcal{M}^{\lambda}\right] \nonumber \\ \\
\label{noise_mf3}                                       & & +\sum_{\lambda}\int_{\mu_{R}}^{\mu_{L}-\hbar\omega}d\varepsilon\textrm{Im}\textrm{Tr}\left[\left(1-2T\right)C\mathcal{M}^{\lambda}A_{L}^{+}\mathcal{M}^{\lambda}\right] \nonumber \\
\end{eqnarray} 
\end{subequations}
and
\begin{subequations}
\begin{eqnarray}
\label{noise_vc1}\frac{\delta S_{\textrm{vc}}}{2e^2/h} &=& -\sum_{\lambda}\int_{\mu_{R}+\hbar\omega}^{\mu_{L}}d\varepsilon2\textrm{Tr}\left[B\mathcal{M}^{\lambda}D^{-}\mathcal{M}^{\lambda}\right] \\
\label{noise_vc2}                                      & & -\sum_{\lambda}\int_{\mu_{R}+\hbar\omega}^{\mu_{L}}d\varepsilon2\textrm{Re}\textrm{Tr}\left[C\mathcal{M}^{\lambda}C^{-}\mathcal{M}^{\lambda}\right] \nonumber \\ \\
\label{noise_vc3}                                      & & -\sum_{\lambda}\int_{\mu_{R}+\hbar\omega}^{\mu_{L}}d\varepsilon2\textrm{Im}\textrm{Tr}\left[C\mathcal{M}^{\lambda}D^{-}\mathcal{M}^{\lambda}\right] \nonumber \\ \\
\label{noise_vc4}                                      & & +\sum_{\lambda}\int_{\mu_{R}}^{\mu_{L}-\hbar\omega}d\varepsilon2\textrm{Im}\textrm{Tr}\left[C\mathcal{M}^{\lambda}D^{+}\mathcal{M}^{\lambda}\right], \nonumber \\
\end{eqnarray} 
\end{subequations}
where $T \equiv A_{R} \Gamma_{L}$. 
Similarly to the current, Eq.~(\ref{qel_noise}) representing the noise correction due to the quasi-elastic current [Eq.~(\ref{qel_current})] does not contribute to steps in the noise signal. 
Thus it will be ignored in our discussion. 

We note that another contribution showing an asymmetric behavior with respect to the applied bias voltage is also ignored~\cite{PRB2005Paulsson, PRB2007Frederiksen_a, PRB2010Haupt, PRB2012Avriller}. 
This asymmetric contribution leads to a logarithmic divergent correction at the threshold bias voltage in the zero temperature limit~\cite{2008PRBEgger, 2009PRBEntin-Wohlman, PRB2010Haupt, PRB2012Avriller}. In fact, it indicates that the lowest-oder perturbation theory is not valid at the threshold bias voltage~\cite{2008PRBEgger, 2009PRBEntin-Wohlman, PRB2010Haupt, PRB2012Avriller}. This divergece might be regularized by introducing damping of vibrational modes, or using a resummation scheme of diagrams~\cite{2009PRBEntin-Wohlman, 2008PRBEgger, PRB2010Haupt, PRB2012Avriller}. 
Away from the threshold bias voltage, the asymmetric term becomes much smaller than the symmetric contribution. 
In addition, it is known that the asymmetric term is negligible for symmetric junctions or for cases where the conductor is close to or far from resonances~\cite{PRL2008Paulsson}. We also remark that the Hartree diagram is not taken into account since it does not lead to step behaviors at the threshold voltage~\cite{PRB2007Frederiksen_a}.

%
%
\begin{figure}[b]
\begin{center}
\includegraphics[width=1.0\columnwidth, clip=true]{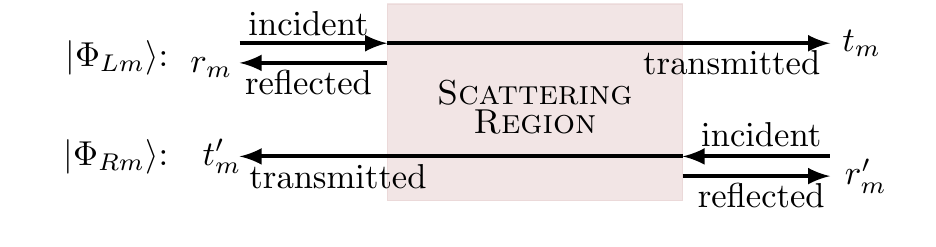} 
\end{center}
\caption{\label{scattering_states} (Color online) Schematic illustration of left-incident and right-incident eigenchannel scattering states. The scattering state consists of incident, transmitted and reflected waves, which are represented by arrows.}  
\end{figure}
\section{Scattering Theory}\label{scattering_theory}
The current and noise corrections based on NEGFs can be expressed in terms of scattering states. In the Landauer-B\"{u}ttiker formalism~\cite{IBMJ1957Landauer, IBMJ1988Buttiker, 1995Datta}, coherent transport is described by scattering states $\{|\Psi_{Lm}\rangle, |\Psi_{Rm}\rangle \}$ and a scattering matrix $\mathbf{S}$ associated with the states. Here indices $\{ m \}$ denote transverse modes called quantum channels~\cite{1995Datta}. The scattering matrix $\mathbf{S}$ is written in the form of a block matrix, 
\begin{equation}
\mathbf{S} = \left(\begin{array}{cc}
\mathbf{S}_{LL} & \mathbf{S}_{LR}\\
\mathbf{S}_{RL} & \mathbf{S}_{RR}
\end{array}\right)=\left(\begin{array}{cc}
\mathbf{r} & \mathbf{t}^{\prime}\\
\mathbf{t} & \mathbf{r}^{\prime}
\end{array}\right),
\end{equation}
where $\mathbf{r}$ and $\mathbf{t}$ ($\mathbf{r}^\prime$ and $\mathbf{t}^\prime$) are reflection and transmission submatrices for left-incident (right-incident) waves.
Note that the energy normalization $\langle \Psi_{\alpha m} (\varepsilon) | \Psi_{\beta n} (\varepsilon^\prime) \rangle = \delta_{\alpha \beta} \delta_{m,n} \delta(\varepsilon - \varepsilon^\prime)$ is used for scattering states~\cite{PRB2007Paulsson}.   
Scattering states $\{|\Psi_{\alpha m} \rangle \}$ are related to Green's functions of the conductor part $G^{r}_{0}$ in the following way~\cite{PRB2009Wang}:
\begin{equation}
\label{identity1} |\Psi_{\alpha m}\rangle = \frac{1}{\sqrt{2\pi}} G_{0}^{r} |W_{\alpha m}\rangle,
\end{equation}
where $|W_{\alpha m} \rangle=\sqrt{2\pi} V_{C\alpha} |u_{\alpha m} \rangle$ and $V_{C\alpha}$ is the coupling Hamiltonian between the conductor and the electrode $\alpha$.
$|u_{\alpha m} \rangle$ is the scattering state when $V_{C\alpha} = 0$, which is just a sum of the incident wave and the totally reflected one. 
$\Gamma_{\alpha}$ is written as $\Gamma_{\alpha} = \sum_{m} |W_{\alpha m} \rangle \langle W_{\alpha m} |$. Then the scattering matrix $\mathcal{S}$~\cite{PRB2009Wang} is 
\begin{eqnarray}
\mathbf{S}_{\alpha m,\beta n} &=&\left(-\delta_{\alpha\beta}\delta_{mn}+i\langle W_{\alpha m}|G_{0}^{r}| W_{\beta n}\rangle\right) \nonumber \\
\label{FisherLee}&=&\left(-\delta_{\alpha\beta}\delta_{mn}+i\sqrt{2\pi}\langle W_{\alpha m}|\Psi_{\beta n}\rangle\right).
\end{eqnarray}

In this work, instead of using the scattering states $\{ |\Psi_{Lm}\rangle, |\Psi_{Rm}\rangle \}$, we choose the transmission eigenchannel representation~\cite{PRB1992Martin, PRB2001Lee, PRB2007Paulsson} $\{ |\Phi_{Lm}\rangle, |\Phi_{Rm}\rangle \}$, in which there is no inter-channel mixing (see Fig.~\ref{scattering_states}). As seen below, el-vib interactions lead to mixing between different eigenchannel states. The transmission eigenchannel representation is achieved by diagonalizing the submatrices $\mathbf{r}$, $\mathbf{t}$, $\mathbf{r}^\prime$, and $\mathbf{t}^\prime$. When the system respects the time-reversal symmetry, it can be done by choosing unitary matrices $U_{L}$ and $U_{R}$ as follows: $\left[U_{L}^{T} \mathbf{r} U_{L} \right]_{mn} = r_{m} \delta_{mn}$, $\left[U_{R}^{T} \mathbf{t} U_{L} \right]_{mn} = t_{m} \delta_{mn}$, $\left[U_{R}^{T} \mathbf{r}^\prime U_{R} \right]_{mn} = r_{m}^\prime \delta_{mn}$, and $\left[U_{L}^{T} \mathbf{t}^\prime U_{R} \right]_{mn} = t_{m}^\prime \delta_{mn}$. See Ref.~\onlinecite{PRB1992Martin,
 PRB2001Lee} for details. Consequently, the original scattering matrix $\mathbf{S}$ is transformed to the scattering matrix $\mathcal{S}$ which is decomposed into a collection of $2\times2$ block scattering matrices $\mathcal{S}_{m}=\left(\begin{smallmatrix} r_{m} & t_{m}^{\prime}\\ t_{m} & r_{m}^{\prime}\end{smallmatrix}\right)$ for eigenchannel states $\{ |\Phi_{Lm}\rangle, |\Phi_{Rm}\rangle \}$. $|\Phi_{\alpha m}\rangle$ is related to $|\Psi_{\alpha m}\rangle$ via $|\Phi_{\alpha m}\rangle = \sum_{n} \left[U_{\alpha} \right]_{mn} |\Psi_{\alpha n}\rangle$.

If the system respects the time-reversal symmetry, one can prove following relations by using Eq. (\ref{identity1}):
\begin{eqnarray}
\label{identity2-1}\hat{\Theta} |\Phi_{Lm}\rangle &=& r_{m}^{*}|\Phi_{Lm}\rangle+t_{m}^{*}|\Phi_{Rm}\rangle\\
\label{identity2-2}\hat{\Theta} |\Phi_{Rm}\rangle &=& t_{m}^{*}|\Phi_{Lm}\rangle+r_{m}^{\prime *}|\Phi_{Rm}\rangle,
\end{eqnarray}
where $\hat{\Theta}$ is the time-reversal operator. Note that these relations are true not only far from the scattering region, but also inside the conductor. 

%
%
\begin{figure}[t]
\begin{center}
\includegraphics[width=1.0\columnwidth, clip=true]{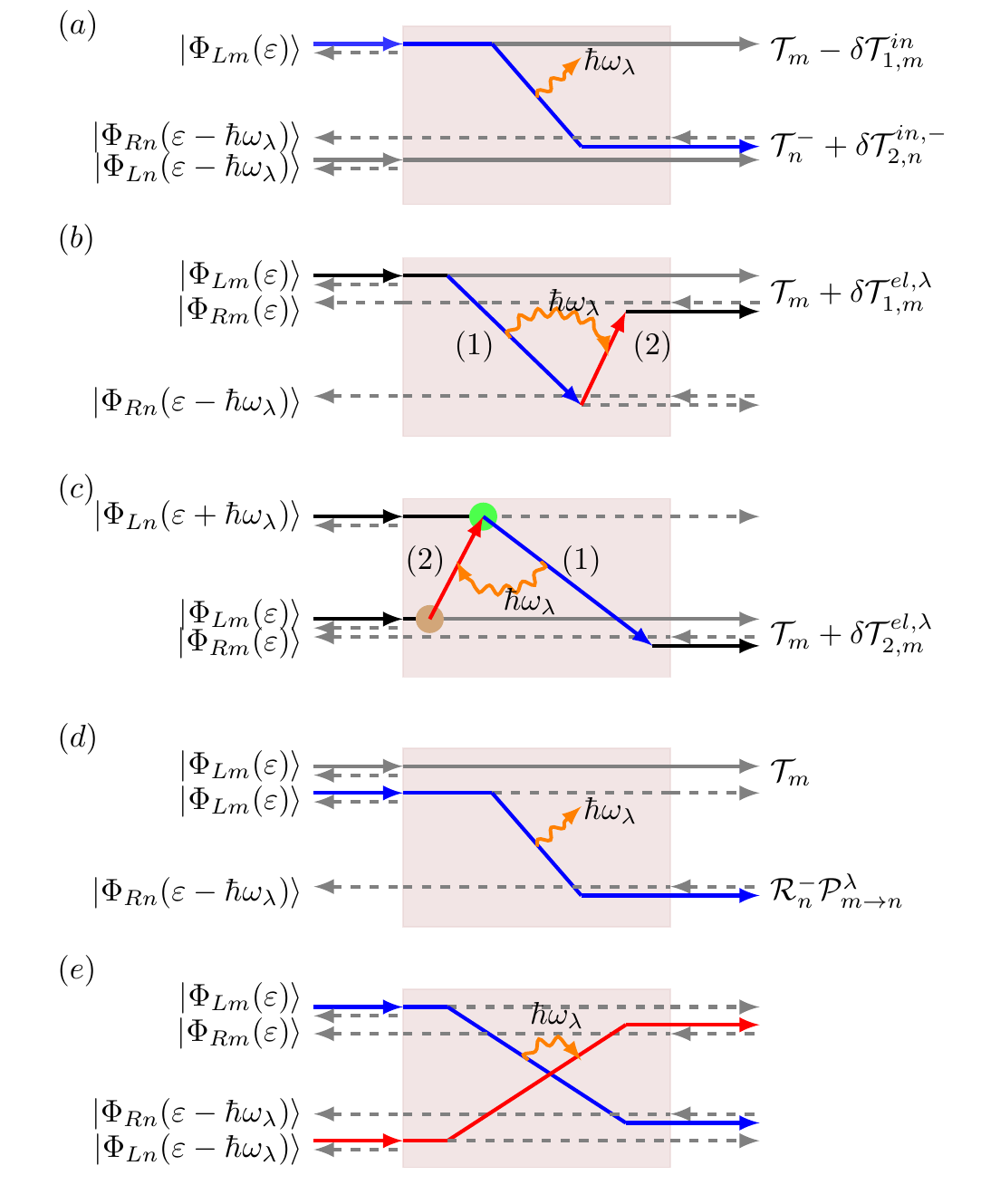} 
\end{center}
\caption{\label{current_noise} (Color online) Schematic explanation of scattering processes contributing to current and noise corrections. 
While solid lines represent parts of the scattering states that electrons follow when they transport to the right electrode, dashed lines are the rest of the scattering states. Blue and red lines indicate vibronic emission and absorption processes, respectively. Gray solid lines represent electron transport without interacting with vibrations, which gives a bare transmission. Orange wiggly lines indicate vibronic energy transfer. On the left side, scattering states and energies corresponding to arrows are indicated. Scattering states at the same energy are vertically shifted for clarity. 
(a) Inelastic contribution to current [Eq.~(\ref{inelastic})]. (b) One-electron scattering process in the elastic contribution to current [Eq.~(\ref{elastic})]. (c) Elastic scattering process involving two electrons (green and brown circles) in Eq.~(\ref{elastic}). (1) and (2) indicate
the order of scattering events. 
Transmission corrections for (a), (b), and (c) scattering processes are indicated on the right side. 
(d) and (e) represent inelastic noise corrections where conducting electrons at two energies differing by $\hbar\omega_{\lambda}$ are correlated. (d) and (e) correspond to Eqs. (\ref{inelastic_noise2}) and (\ref{elastic_noise2}), respectively.
Note that (a) and (d) correspond to $|\Psi_{1}\rangle = |\psi_{1}\rangle_{e} \otimes |1\rangle_{v}$, while (b), (c), and (e) describe $|\psi_{2}^{(1)}\rangle_{e} \otimes |0\rangle_{v}$ of $|\Psi_{2}\rangle$ in Sec.~\ref{general_picture}. 
}  
\end{figure}
\subsection{Current Corrections}\label{Subsec_Current}
In Ref.~\onlinecite{PRB2013SKim}, using Eqs. (\ref{identity1})-(\ref{identity2-2}), we obtained the current correction $\delta I$ in terms of scattering states:
\begin{equation}
\delta I = \delta I_{\textrm{1BA}} + I_{\textrm{2BA}},
\end{equation}
where 
\begin{eqnarray}
\label{inelastic} \delta I_{\textrm{1BA}} &=& \frac{2e}{h} \left(2\pi\right)^{2}\sum_{\lambda}\sum_{m,n}\int_{\mu_{R}+\hbar\omega_{\lambda}}^{\mu_{L}}d\varepsilon \nonumber \\
                                          & &  \left(\mathcal{R}_{n}^{-}-\mathcal{T}_{m}\right) \left|\langle\Phi_{Rn}^{-}|\mathcal{M}^{\lambda}|\Phi_{Lm}\rangle\right|^{2},
\end{eqnarray}
and
\begin{eqnarray}
\label{elastic} \delta I_{\textrm{2BA}} &=& -\frac{2e}{h}\left(2\pi\right)^{2}\sum_{\lambda}\sum_{m,n}\int_{\mu_{R}+\hbar\omega_{\lambda}}^{\mu_{L}}d\varepsilon \nonumber \\
                                        & &  \textrm{Re}\left[  r_{m}^{\prime}t_{m}^{*}\langle\Phi_{Rm}|\mathcal{M}^{\lambda}|\Phi_{Rn}^{-}\rangle\langle\Phi_{Rn}^{-}|\mathcal{M}^{\lambda}|\Phi_{Lm}\rangle\right]  \nonumber\\
                                        & &  +\frac{2e}{h}\left(2\pi\right)^{2}\sum_{\lambda}\sum_{m,n}\int_{\mu_{R}}^{\mu_{L}-\hbar\omega_{\lambda}}d\varepsilon \nonumber \\
                                        & &  \textrm{Re}\left[r_{m}^{\prime}t_{m}^{*}\langle\Phi_{Rm}|\mathcal{M}^{\lambda}|\Phi_{Ln}^{+}\rangle\langle\Phi_{Ln}^{+}|\mathcal{M}^{\lambda}|\Phi_{Lm}\rangle\right], \nonumber \\
\end{eqnarray}
where $\mathcal{R}^{-}_{n} = \left|r_{n} \left( \varepsilon-\hbar\omega_{\lambda} \right) \right|^2$ and $\mathcal{T}_{m} = \left|t_{m} \left( \varepsilon \right) \right|^2$.
Equation (\ref{inelastic}) represents the inelastic scattering process where a conducting electron is scattered off by emitting one vibron $\hbar\omega_{\lambda}$ (see Fig.~\ref{current_noise}(a)). In contrast, Eq.~(\ref{elastic}) describes the elastic scattering contribution to the current with a vibrational emission-reabsorption process. The first term of Eq.~(\ref{elastic}) is an one-electron scattering process where only one electron is involved in the vibrational emission-reabsorption process as depicted in Fig.~\ref{current_noise}(b). In this case, energy of each conducting electron does not change. 
The second term of Eq.~(\ref{elastic}) involves successive scatterings of two electrons with local vibrons (see Fig.~\ref{current_noise}(c)). One vibrational energy quantum $\hbar \omega_{\lambda}$ is transferred from one electron to the other via the vibrational emission-reabsorption process. While energies of the two electrons involved in the two-electron process change (one increases and the other decreases), the total energy of conducting electrons does not change, so it is regarded as the elastic process. In fact, Eqs. (\ref{inelastic}) and (\ref{elastic}) correspond to the first and second Born approximations in a standard scattering theory~\cite{PRB1970Davis}. See Ref.~\onlinecite{PRB2013SKim} for this specification based on charge transfers between scattering states.

For the inelastic process, the right scattering state $|\Phi_{Rn} (\varepsilon - \hbar \omega_{\lambda}) \rangle$ should be empty, when the electron is incident at $\varepsilon$ from the left side. 
Otherwise, the inelastic scattering is prohibited due to the Pauli exclusion principle. The inelastic scattering, therefore, can occur when $\varepsilon \in [\mu_R+\hbar\omega_{\lambda},\mu_L]$. Two elastic scattering processes take place at different energy ranges. 
Let us consider elastic corrections at $\varepsilon$. Similar to the inelastic process, the one-electron process requires an empty right scattering state $|\Phi_{Rn} (\varepsilon - \hbar \omega_{\lambda}) \rangle$ that a left-incident electron at $\varepsilon$ can occupy after emitting one vibron. In contrast, the two-electron process assumes a left-incident electron occupying $|\Phi_{Ln}(\varepsilon+\hbar\omega_{\lambda})\rangle$, which is first scattered off to an empty right scattering state $|\Phi_{Rn} (\varepsilon) \rangle$~\cite{PRB2013SKim}. Considering these, one can realize that the one-electron and two-electron elastic processes can occur at $\varepsilon \in [\mu_R+\hbar\omega_{\lambda},\mu_L]$ and $\varepsilon \in [\mu_R,\mu_L-\hbar\omega_{\lambda}]$ respectively, as shown in Eq.~(\ref{elastic}). 

These scattering processes can be written as transmission corrections to a bare transmission $\mathcal{T}_{m}(\varepsilon)$ when there is no el-vib interaction. Considering scattering with the $\lambda$th vibrational mode, the inelastic scattering process [Eq.~(\ref{inelastic})] gives two transmission corrections at $\varepsilon$ and $\varepsilon-\hbar\omega_{\lambda}$:
\begin{eqnarray}
\label{inelasticT1} \delta \mathcal T_{1,m}^{\textrm{in},\lambda}(\varepsilon) &=& - \mathcal{T}_{m} \sum_{n} \mathcal{P}^{\lambda}_{m \rightarrow n} \mathcal{F}(\varepsilon)\\
\label{inelasticT2} \delta \mathcal T_{2,n}^{\textrm{in},\lambda}(\varepsilon-\hbar\omega_{\lambda}) &=& \sum_{m} \mathcal{R}_{n}^{-} \mathcal{P}^{\lambda}_{m \rightarrow n} \mathcal{F}(\varepsilon),
\end{eqnarray}
where 
\begin{eqnarray}
\mathcal{P}^{\lambda}_{m \rightarrow n} &=& (2\pi)^2 |\langle\Phi_{Rn}^{-}|\mathcal{M}^{\lambda}|\Phi_{Lm}\rangle|^{2} \\
\mathcal{F}(\varepsilon) &=& \theta (\varepsilon - \mu_R - \hbar\omega_{\lambda}) [1-\theta(\varepsilon - \mu_L)].
\end{eqnarray}
Here $\theta(x)$ denotes the step function. $\mathcal{F}(\varepsilon)$ accounts for the energy window in which the inelastic scattering is possible as discussed above. 
When interacting with the $\lambda$th vibrational mode, the electron at the left scattering state $|\Phi_{Lm}(\varepsilon)\rangle$ is inelastically scattered off to the right scattering state $|\Phi_{Rn}(\varepsilon-\hbar\omega_{\lambda})\rangle$ 
with the scattering probability $\mathcal{P}^{\lambda}_{m \rightarrow n}$. 
In comparison with the case where there is no el-vib interaction, the transmission of the $m$th eigenchannel at $\varepsilon$ is reduced
by Eq.~(\ref{inelasticT1}) in which the total scattering probability is taken into account by summing over all right-incident eigenchannels.
For electrons scattered to the right scattering state $|\Phi_{Rn}(\varepsilon-\hbar\omega_{\lambda})\rangle$, 
the transmission probability to move to the right electrode is given by the reflection amplitude $\mathcal{R}_{n}(\varepsilon-\hbar\omega_{\lambda})$. 
Thus, by considering the scattering probability $\mathcal{P}^{\lambda}_{m \rightarrow n}$ together, the transmission correction of scattered electrons from the $m$th eigenchannel to the $n$th eigenchannel is given by $\mathcal{R}_{n}^{-} \mathcal{P}_{m \rightarrow n}^{\lambda}$.
Summing over all incident states $|\Phi_{Lm}(\varepsilon)\rangle$, the total transmission correction to the eigenchannel $n$ is given by Eq.~(\ref{inelasticT2}). In contrast, the elastic scattering process gives corrections to the transmission at $\varepsilon$,
\begin{equation}
\label{elasticT} \delta \mathcal T^{\textrm{el},\lambda}_{m}(\varepsilon) = \delta \mathcal T^{\textrm{el},\lambda}_{1,m}(\varepsilon) + \delta \mathcal T^{\textrm{el},\lambda}_{2,m}(\varepsilon)
\end{equation}
where
\begin{widetext}
\begin{eqnarray}
\label{elasticT1}
\delta \mathcal T^{\textrm{el},\lambda}_{1,m}(\varepsilon) &=& -(2\pi)^2 \sum_{n}\textrm{Re}\left[  r_{m}^{\prime}t_{m}^{*}\langle\Phi_{Rm}|\mathcal{M}^{\lambda}|\Phi_{Rn}^{-}\rangle\langle\Phi_{Rn}^{-}|\mathcal{M}^{\lambda}|\Phi_{Lm}\rangle\right]\mathcal{F} (\varepsilon) \\
\label{elasticT2}
\delta \mathcal T^{\textrm{el},\lambda}_{2,m}(\varepsilon) &=&  (2\pi)^2 \sum_{n}\textrm{Re}\left[r_{m}^{\prime}t_{m}^{*}\langle\Phi_{Rm}|\mathcal{M}^{\lambda}|\Phi_{Ln}^{+}\rangle\langle\Phi_{Ln}^{+}|\mathcal{M}^{\lambda}|\Phi_{Lm}\rangle\right]\mathcal{F} (\varepsilon + \hbar\omega_{\lambda}). 
\end{eqnarray}
\end{widetext}
Equations~(\ref{elasticT1}) and (\ref{elasticT2}) corresponds to corrections due to one-electron and two-electron elastic processes respectively. 
In terms of these transmission corrections, the current corrections $\delta I_{\textrm{1BA}}$ and $\delta I_{\textrm{2BA}}$ are written as 
\begin{eqnarray}
\delta I_{\textrm{1BA}} &=& \frac{2e}{h}\sum_{\lambda,m} \int d\varepsilon \delta \mathcal{T}_{1,m}^{\textrm{in}, \lambda}(\varepsilon) + \delta \mathcal{T}^{\textrm{in}, \lambda}_{2,n}(\varepsilon-\hbar\omega_{\lambda}) \\
\delta I_{\textrm{2BA}} &=& \frac{2e}{h}\sum_{\lambda,m} \int d\varepsilon \delta \mathcal T^{\textrm{el},\lambda}_{1,m}(\varepsilon) + \delta \mathcal T^{\textrm{el},\lambda}_{2,m}(\varepsilon).
\end{eqnarray}

%
%
\subsection{Noise Corrections}\label{Subsec_Noise}
Using Eqs.~(\ref{identity1})-(\ref{identity2-2}), one can obtain the scattering state description of the noise correction:
\begin{equation}
\delta S = \delta S^{\textrm{in}}_{1} + S^{\textrm{in}}_{2} + S^{\textrm{el}}_{1} + S^{\textrm{el}}_{2},
\end{equation}
where
\begin{widetext}
\begin{eqnarray}
\frac{\delta S^{\textrm{in}}_{1}}{2e^{2}/h} &=& -\left(2\pi\right)^{2}\sum_{\lambda,m,n}\int_{\mu_{R}+\hbar\omega_\lambda}^{\mu_{L}}d\varepsilon\left(1-2\mathcal{T}_{m}\right)\mathcal{T}_{m}\left|\langle\Phi_{Rn}^{-}|\mathcal{M}^{\lambda}|\Phi_{Lm}\rangle\right|^{2} \nonumber \\
\label{inelastic_noise1}                         & & +\left(2\pi\right)^{2}\sum_{\lambda,m,n}\int_{\mu_{R}+\hbar\omega_\lambda}^{\mu_{L}}d\varepsilon\left(1-2\mathcal{T}_{n}^{-}\right)\mathcal{R}_{n}^{-}\left|\langle\Phi_{Rn}^{-}|\mathcal{M}^{\lambda}|\Phi_{Lm}\rangle\right|^{2}
\end{eqnarray}

\begin{eqnarray}
\label{inelastic_noise2}\frac{\delta S^{\textrm{in}}_{2}}{2e^{2}/h} &=& -2\left(2\pi\right)^{2}\sum_{\lambda,m,n}\int_{\mu_{R}+\hbar\omega_\lambda}^{\mu_{L}}d\varepsilon\mathcal{T}_{m}\mathcal{R}_{n}^{-}\left|\langle\Phi_{Rn}^{-}|\mathcal{M}^{\lambda}|\Phi_{Lm}\rangle\right|^{2}
\end{eqnarray} 

\begin{eqnarray}
\label{elastic_noise1}\frac{\delta S^{\textrm{el}}_{1}}{2e^{2}/h} &=& -\left(2\pi\right)^{2}\sum_{\lambda,m,n}\int_{\mu_{R}+\hbar\omega_\lambda}^{\mu_{L}}d\varepsilon\left(1-2\mathcal{T}_{m}\right)\textrm{Re}\left[r_{m}^{\prime}t_{m}^{*}\langle\Phi_{Rm}|\mathcal{M}^{\lambda}|\Phi_{Rn}^{-}\rangle\langle\Phi_{Rn}^{-}|\mathcal{M}^{\lambda}|\Phi_{Lm}\rangle\right] \nonumber \\
                                                                  & & +\left(2\pi\right)^{2}\sum_{\lambda,m,n}\int_{\mu_{R}}^{\mu_{L}-\hbar\omega_\lambda}d\varepsilon\left(1-2\mathcal{T}_{m}\right)\textrm{Re}\left[r_{m}^{\prime}t_{m}^{*}\langle\Phi_{Rm}|\mathcal{M}^{\lambda}|\Phi_{Ln}^{+}\rangle\langle\Phi_{Ln}^{+}|\mathcal{M}^{\lambda}|\Phi_{Lm}\rangle\right]
\end{eqnarray} 

\begin{eqnarray}
\label{elastic_noise2}\frac{\delta S^{\textrm{el}}_{2}}{2e^{2}/h} &=& 2\left(2\pi\right)^{2}\sum_{\lambda,m,n}\int_{\mu_{R}+\hbar\omega_\lambda}^{\mu_{L}}d\varepsilon\textrm{Re}\left\{ r_{m}^{\prime}t_{m}^{*}r_{n}^{\prime-}t_{n}^{-*}\langle\Phi_{Rm}|\mathcal{M}^{\lambda}|\Phi_{Ln}^{-}\rangle\langle\Phi_{Rn}^{-}|\mathcal{M}^{\lambda}|\Phi_{Lm}\rangle\right\}.
\end{eqnarray}  
\end{widetext}
Here Eqs.~(\ref{inelastic_noise1}) and (\ref{inelastic_noise2}) are interpreted as inelastic processes, while Eqs.~(\ref{elastic_noise1}) and (\ref{elastic_noise2}) are elastic ones. We first consider Eqs.~(\ref{inelastic_noise1}) and (\ref{elastic_noise1}). They can be understood as current correlations of electrons at one energy $\varepsilon$. When there is no el-vib interaction, the shot noise at zero temperature is written as 
\begin{equation}
\frac{S_{0}}{2e^2/h} = \sum_{m}\int^{\mu_{L}}_{\mu_{R}} d\varepsilon \mathcal{T}_{m} \left( 1 - \mathcal{T}_{m} \right).
\end{equation}
If the transmission $\mathcal{T}_{m}(\varepsilon)$ is slightly changed by $\delta \mathcal{T}_{m}$, {\it i.e.}, $\mathcal{T}_{m} \rightarrow \mathcal{T}_{m} + \delta \mathcal{T}_{m}$,
the shot noise is corrected as $S = S_{0} + \delta S$, where
\begin{equation}
\label{noise_correction}\frac{\delta S}{2e^2/h} = \sum_{m}\int^{\mu_{L}}_{\mu_{R}} d\varepsilon \left( 1 - 2 \mathcal{T}_{m} \right) \delta \mathcal{T}_{m} + O(\delta \mathcal{T}^2).
\end{equation}
In the previous subsection, it is discussed that the inelastic scattering process leads to transmission corrections $\delta \mathcal{T}^{\textrm{in},\lambda}_{1,m}(\varepsilon)$ [Eq.~(\ref{inelasticT1})] and $\delta \mathcal{T}^{\textrm{in},\lambda}_{2,n}(\varepsilon-\hbar\omega_{\lambda})$ [Eq.~(\ref{inelasticT2})]
to bare transmissions $\mathcal{T}_{m}(\varepsilon)$ and $\mathcal{T}_{n}(\varepsilon-\hbar\omega_{\lambda})$, respectively. For the elastic contribution that consists of one-electron and two-electron processes as discussed before, $\delta \mathcal{T}^{\textrm{el},\lambda}_{m}$ [Eq.~(\ref{elasticT})] is added to a bare transmission $\mathcal{T}_{m}(\varepsilon)$.
Thus, when these transmission corrections $\delta \mathcal{T}^{\textrm{in},\lambda}_{1,m}(\varepsilon)$, $\delta \mathcal{T}^{\textrm{in},\lambda}_{2,n}(\varepsilon-\hbar\omega_{\lambda})$, and $\delta \mathcal{T}^{\textrm{el},\lambda}_{m}$ are plugged into Eq.~(\ref{noise_correction}).
one can check that Eqs.~(\ref{inelastic_noise1}) and (\ref{elastic_noise1}) are recovered. 
Further, Eqs.~(\ref{inelastic_noise1}) and (\ref{elastic_noise1}) are identified as inelastic and elastic corrections to the shot noise respectively, considering that they are derived from the transmission corrections due to inelastic and elastic scattering processes.  

In contrast, Eqs.~(\ref{inelastic_noise2}) and (\ref{elastic_noise2}) accounts for current correlations of electrons at two different energies $\varepsilon$ and $\varepsilon-\hbar\omega_{\lambda}$.
To be specific, Eq.~(\ref{inelastic_noise2}) describes the current correlation between an electron of $|\Phi_{Lm}(\varepsilon)\rangle$ that is not scattered by vibrons, and another electron that initially occupies $|\Phi_{Lm}(\varepsilon)\rangle$, but is transferred to $|\Phi_{Rn}(\varepsilon-\hbar\omega_{\lambda})\rangle$ by emitting one vibron (see Fig.~\ref{current_noise}(d)). On the contrary, Eq.~(\ref{elastic_noise2}) is the current correlation of two electrons exchanging a vibronic energy $\hbar\omega_{\lambda}$ (see Fig.~\ref{current_noise}(e)). 
The electron of $|\Phi_{Lm}(\varepsilon)\rangle$ is first scattered off to the right scattering state $|\Phi_{Rn}(\varepsilon-\hbar\omega_{\lambda})\rangle$ by emitting one vibron. 
After that, another electron of $|\Phi_{Ln}(\varepsilon-\hbar\omega_{\lambda})\rangle$ absorbs the vibron, thereby being excited to $|\Phi_{Rm}(\varepsilon)\rangle$. Considering whether the total electronic energy is conserved or not, it is obvious that Eqs.~(\ref{inelastic_noise1}) and (\ref{elastic_noise2}) are inelastic and elastic current correlations, respectively. Note that Eq.~(\ref{elastic_noise2}) is written in the interference-like form of Eq.~(\ref{elastic_general}), which is identified as elastic in Sec.~\ref{general_picture}.

We remark how Eqs.~(\ref{inelastic_noise1})-(\ref{elastic_noise2}) are related to the mean-field correction $\delta S_{\textrm{mf}}$ and the vertex correction $\delta S_{\textrm{vc}}$~\cite{PRB2010Haupt}. Inelastic corrections Eqs.~(\ref{inelastic_noise1}) and (\ref{inelastic_noise2}) are derived from both $\delta S_{\textrm{mf}}$ and $\delta S_{\textrm{vc}}$. While Eq.~(\ref{elastic_noise1}) arises solely from the mean-field correction $\delta S_{\textrm{mf}}$, Eq.~(\ref{elastic_noise2}) is derived from the vertex correction $\delta S_{\textrm{vc}}$. See Appedix~\ref{derivation} for detailed derivations. 

This scattering-state description enables to quantitatively calculate intra-channel and inter-channel scattering contributions to the inelastic signal as well as elastic and inelastic ones. It can be complementary to the existing analysis based on scattering rates~\cite{PRL2008Paulsson} that helps qualitatively figure out what scattering process is dominant in the inelastic signal, which is known as the propensity rule~\cite{PRL2008Paulsson, PRB2007Gagliardi, JChemPhys2006Troisi}.  For example, it is possible that some vibrational modes, which are minor in the conductance steps, can give visible contributions to the inelastic noise signals, as reported in Ref.~\onlinecite{PRB2012Avriller}. 
Different visibilities of those modes in the conductance and the noise signal are not understood by calculating scattering rates, which are in the form of Fermi's golden rule~\cite{PRL2008Paulsson}. Instead, our description can give a quantitative explanation to those visibilities by clarifying the interplay of intra-channel and inter-channel scattering contributions, or that of elastic and inelastic scattering processes. See Appendix~\ref{multichannel_systems} for further discussions.
%
%

\section{Discussions}\label{discussion}
\begin{figure}[t]
\begin{center}
\includegraphics[width=1.0\columnwidth, clip=true]{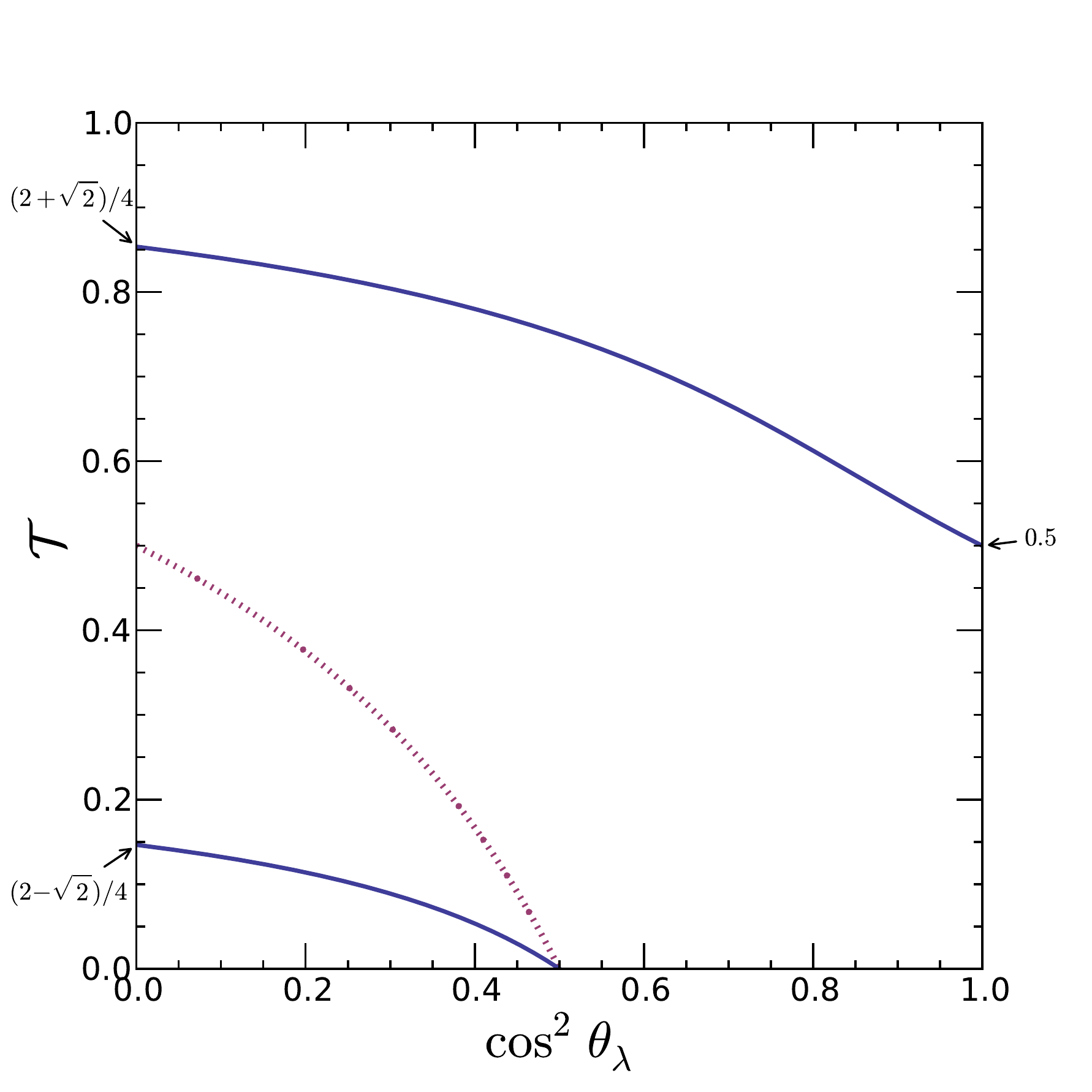} 
\end{center}
\caption{\label{fig_crossover} (Color online) Crossover transmissions for the conductance and the shot noise. Purple dotted line and blue solid line represent possible crossover transmissions of conductance and noise corrections, respectively. }
\end{figure}
We apply our scattering-theory description of the current and noise corrections to single-channel systems. 
We further adopt the approximation known as the {\it extended wide-band limit} (EWBL), in which Green's functions and coupling functions are replaced by those at Fermi energy $\varepsilon_{F}$: $G^{r}_{0}(\varepsilon) \approx G^{r}_{0}(\varepsilon_{F})$ and $\Gamma_{\alpha}(\varepsilon) \approx \Gamma_{\alpha}(\varepsilon_{F})$~\cite{PRB2005Paulsson, PRB2007Frederiksen_a}.
In this approximation, scattering states and their scattering matrices are replaced by those at $\varepsilon_{F}$. 
The EWBL can be valid when the density of states of the system is slowly varying over a few vibrational energies around $\varepsilon_{F}$. 

In Ref.~\onlinecite{PRB2013SKim}, we show that the conductance change at the threshold voltage $eV=\hbar\omega_{\lambda}$ is written as
\begin{equation}
\label{conductance_single_channel}
\frac{\Delta G_{\lambda}}{2e^2/h} = (2\pi)^2 \left(1 - 2\mathcal{T} - 2\mathcal{R} \cos^2 \theta_{\lambda} \right) \left| \mathbb{M}^{\lambda}_{RL} \right|^2,
\end{equation}
where 
$\mathbb{M}^{\lambda}_{\alpha \beta}=\langle \Phi_{\alpha} | \mathcal{M}^{\lambda} | \Phi_{\beta} \rangle$ 
and $\theta_{\lambda} = \arg \left[r^{\prime} t^{*} \mathbb{M}^{\lambda}_{R L} \right]$.
By solving $\Delta G_{\lambda} = 0$, it is shown that the crossover transmission $\mathcal{T}^{c}_{\textrm{cr}}$ for the conductance satisfies
the following relation:
\begin{equation}
\label{crossover}
\mathcal{T} = \frac{1-2\cos^2 \theta_{\lambda}}{2\left(1-\cos^2 \theta_{\lambda} \right)}.
\end{equation}
Since $\cos^2 \theta_{\lambda}$ generally depends on $\mathcal{T}$ (and other system parameters), Eq.~(\ref{crossover}) does not provide the analytic expression of the crossover transmission $\mathcal{T}^{c}_{\textrm{cr}}$.
One can nonetheless find out a possible range of the crossover transmission $\mathcal{T}^{c}_{\textrm{cr}}$ from Eq.~(\ref{crossover}). 
Considering $0 \leq \cos^2 \theta_{\lambda} \leq 1$, it can be shown that $\mathcal{T}^{c}_{\textrm{cr}} \leq 0.5$, where the equality holds when $\cos^2 \theta_{\lambda}=0$.
When $\cos^2 \theta_{\lambda} \geq 0.5$, $\mathcal{T}$ becomes negative, and it implies that there is no crossover (see Fig.~\ref{fig_crossover})~\cite{PRB2013SKim}.


Following Ref.~\onlinecite{PRB2010Haupt}, the {\it inelastic noise signal} is defined as 
\begin{equation}
\Delta S_{\lambda}^\prime \equiv \left.\frac{d \delta S}{dV}\right|_{\hbar\omega_{\lambda}+\eta} - \left.\frac{d \delta S}{dV}\right|_{\hbar\omega_{\lambda}-\eta},
\end{equation}
where $\eta$ stands for a small positive value.
The inelastic noise signal in the EWBL is 
\begin{eqnarray}
\label{noise_single_channel}
\frac{\Delta S_{\lambda}^{\prime}}{2e^3/h} &=& (2\pi)^2 \left[\left(8-8\cos^2 \theta_{\lambda} \right)\mathcal{T}^2 \right. \nonumber \\
                                                  && +\left(10\cos^2\theta_{\lambda}-8\right)\mathcal{T} \nonumber \\
                                                  && \left. +\left(1-2\cos^2 \theta_{\lambda} \right) \right] \left| \mathbb{M}^{\lambda}_{RL} \right|^2.
\end{eqnarray}
$\Delta S^{\prime}_{\lambda}$ is also decomposed into inelastic and elastic scattering contributions, $\Delta S^{\prime}_{\lambda} = \Delta S^{\textrm{in}\prime}_{\lambda}+\Delta S^{\textrm{el}\prime}_{\lambda}$, where
\begin{eqnarray}
\label{inelastic_noise_signal}\frac{\Delta S_{\lambda}^{\textrm{in}\prime}}{2e^3/h} &=& (2\pi)^2 \left( 6\mathcal{T}^2 - 6 \mathcal{T} +1 \right) \left| \mathbb{M}^{\lambda}_{RL} \right|^2 \\
\label{elastic_noise_signal}\frac{\Delta S_{\lambda}^{\textrm{el}\prime}}{2e^3/h} &=& (2\pi)^2  \left[2(2\mathcal{T}-1)\mathcal{R}\cos^2\theta_{\lambda} \right. \nonumber\\
&& \left. + 2\mathcal{R}\mathcal{T}\cos 2\theta_{\lambda} \right]\left| \mathbb{M}^{\lambda}_{RL} \right|^2.
\end{eqnarray}
Here Eqs.~(\ref{inelastic_noise1}) and (\ref{inelastic_noise2}) [Eqs.~(\ref{elastic_noise1}) and (\ref{elastic_noise2})] give rise to $\Delta S_{\lambda}^{\textrm{in}\prime}$ [$\Delta S_{\lambda}^{\textrm{el}\prime}$]. 
While $\Delta S^{\textrm{in}\prime}_{\lambda}$ is determined only by $\mathcal{T}$, 
$\Delta S^{\textrm{el}\prime}_{\lambda}$ depends on the phase information of the system via $\theta_{\lambda}$, as discussed in Sec.~\ref{general_picture}. In contrast to the conductance variation where the elastic contribution is always negative, the elastic term $\Delta S^{\textrm{el}\prime}_{\lambda}$ can be either positive or negative. From Eq.~(\ref{noise_single_channel}), it is readily shown that the crossover transmission $\mathcal{T}_{\textrm{cr}}^{n}$ for the inelastic noise signal satisfies
\begin{equation}
\label{crossover_noise}
\mathcal{T} = \frac{\left(4-5\cos^2\theta_{\lambda}\right)\pm\sqrt{9\cos^4\theta_{\lambda}-16\cos^2\theta_{\lambda}+8}}{8\left(1-\cos^2\theta_{\lambda}\right)}.
\end{equation}
We plot two solutions of Eq.~(\ref{crossover_noise}) (blue solid lines) in Fig.~\ref{fig_crossover}. 
As shown, the crossover for the noise signal can occur between $0$ and $(2-\sqrt{2})/4$ or between $0.5$ and $(2+\sqrt{2})/4$. The crossover transmission $\mathcal{T}_{\textrm{cr}}^{n}$ cannot exceed $(2+\sqrt{2})/4$. 

\subsection{Mirror Symmetry}
\begin{figure}[t]
\begin{center}
\includegraphics[width=1.0\columnwidth, clip=true]{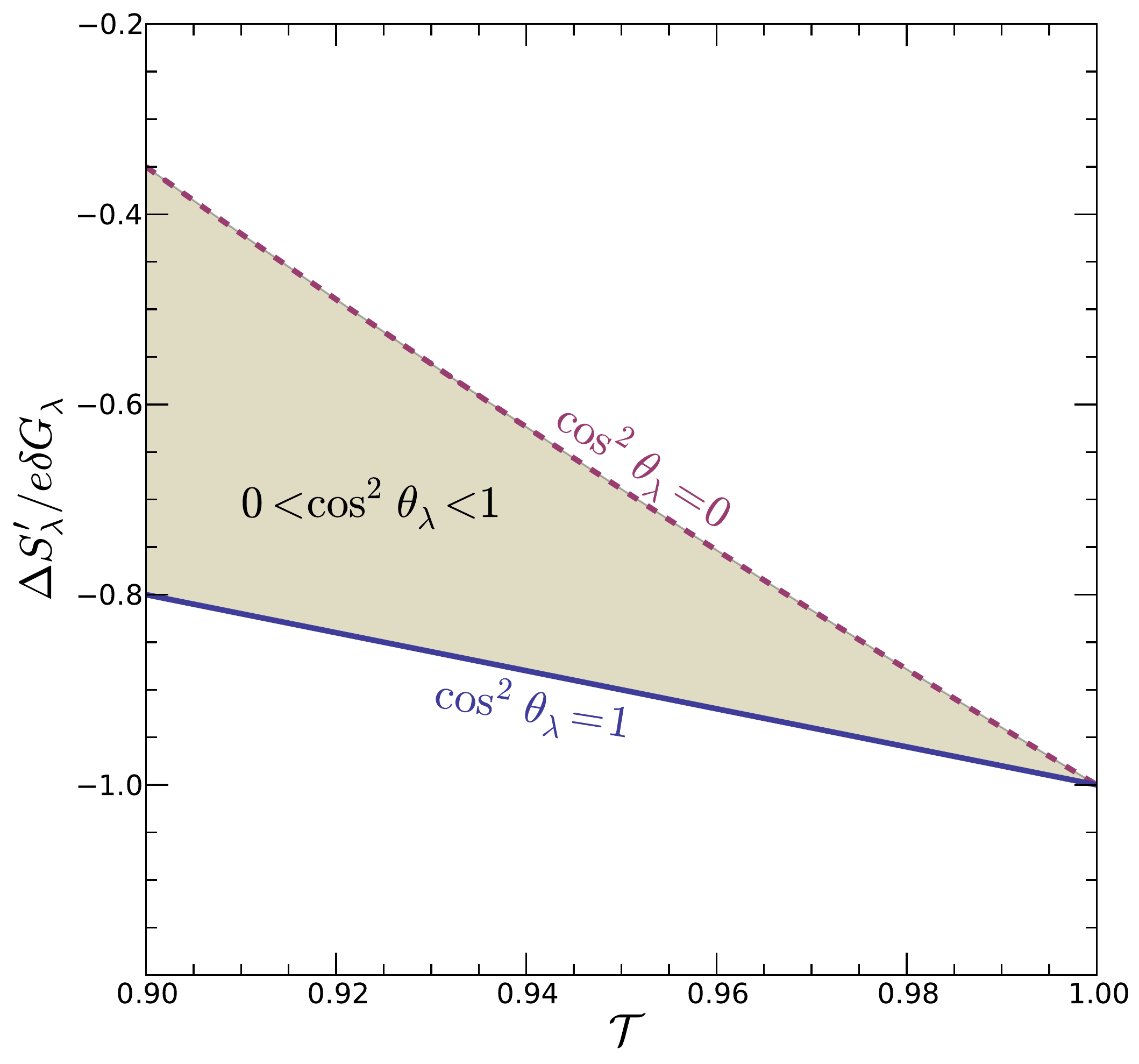} 
\end{center}
\caption{\label{fig_ratio} (Color online) The ratio $\Delta S^{\prime}_{\lambda} / e \delta G_{\lambda}$ in a high transmission regime.
Purple dashed line and blue solid line indicate the ratios $\Delta S^{\prime}_{\lambda} / e \delta G_{\lambda}$ of $\cos^{2}\theta_{\lambda}=0$ and $\cos^{2}\theta_{\lambda}=1$, respectively. When $\cos^{2} \theta_{\lambda} < 1$, the ratio is located in the green region.}
\end{figure}
Now we focus on systems with the mirror-reflection symmetry along the transport direction. 
Mirror-symmetric systems are of particular interest because bridge junctions in many experiments roughly respect this symmetry. As discussed in Ref.~\onlinecite{PRB2013SKim}, one can analytically calculate the crossover transmissions since $\cos^{2} \theta_{\lambda}$ is just a number for mirror-symmetric systems.
For mirror-symmetric systems, vibrational modes are either even or odd mirror-symmetric.
When $\mathbf{R}$ denotes the mirror-reflection operator, even and odd mirror-symmetric modes satisfy the following relations~\cite{PRB2013SKim}:
\begin{eqnarray}
\mathbf{R} \mathcal{M}^{\lambda}_{\textrm{even}} \mathbf{R}^{\dagger} &=& \mathcal{M}^{\lambda}_{\textrm{even}} \\
\mathbf{R} \mathcal{M}^{\lambda}_{\textrm{odd}} \mathbf{R}^{\dagger} &=& -\mathcal{M}^{\lambda}_{\textrm{odd}}.
\end{eqnarray}
Moreover, left-incident and right-incident scattering states are related to each other as $|\Phi_{L,R}\rangle = \mathbf{R} |\Phi_{R,L}\rangle$~\cite{PRB2013SKim, PRB2001Lee}.
Using these relations, it is found that $\cos^2 \theta_{\lambda}=0$ $(1)$ for even (odd) modes~\cite{PRB2013SKim}. 
It means that the elastic contribution to the conductance variation, which is $-2\mathcal{R}\cos^2 \theta_{\lambda}$ in Eq.~(\ref{conductance_single_channel}), vanishes for even vibrational modes, while it becomes $-2\mathcal{R}$ for odd modes. 
From Eq.~(\ref{crossover}), it is shown that even vibrational modes exhibit the crossover of the conductance step at $\mathcal{T}=0.5$, while odd modes do not show any crossover in the conductance step (see  Fig.~\ref{fig_crossover}). 

The inelastic noise signal also exhibits different crossover behaviors, depending on whether vibrational modes are even or odd. For even vibrational modes $(\cos^2 \theta_{\lambda}=0$), the inelastic noise signal changes its sign at $\mathcal{T} = (2\pm\sqrt{2})/4$.
In contrast, odd vibrational modes $(\cos^2 \theta_{\lambda}=1$) have only one transition between positive and negative noise signals at $\mathcal{T}=0.5$. 
This result is consistent with the fact that sign changes in the inelastic noise signal take place at  $\mathcal{T} = (2\pm\sqrt{2})/4$ in the single-level model~\cite{PRL2012Kumar} and the two-site model interacting with the even-parity mode~\cite{PRB2012Avriller}, both of which are symmetrically coupled to electrodes. 
While the aforementioned models deal only with even modes, the parity effect on the crossover behavior of the noise signal is observed in tight-binding models of $N$-site atomic chains~\cite{PRB2012Avriller}.
For chains with an even number of sites $(N=2p)$ where even vibrational modes dominantly contribute to conductance and noise corrections, it is reported that the noise signal crossover occurs at $\mathcal{T}=(2\pm\sqrt{2})/4$. 
By contrast, the crossover of the noise signal at $\mathcal{T}=0.5$ is observed for chains of an odd number of sites $(N=2p+1)$ where odd modes are dominant in conductance and noise corrections.
Our theory shows that the crossover behaviors reported in the previous studies~\cite{PRL2012Kumar, PRB2012Avriller} are not coincident, but they are generally expected in any single-channel system respecting the mirror symmetry. 

Reference~\onlinecite{PRB2012Avriller} also performed DFT calculations on gold atomic chains varying in lengths and configurations. The computed ratios of $\Delta S^{\prime}_{\lambda}$ to $e \Delta G_{\lambda}$ for the atomic chains are compared with the ratio predicted by the two-site model~\cite{PRB2012Avriller}. The calculated ratios do not exactly match that of the two-site analytic model. Rather, it seems that the ratio of the analytic model might be an upper bound to the ratios calculated from DFT results. From Eqs.~(\ref{conductance_single_channel}) and (\ref{noise_single_channel}), the ratio of $\Delta S^{\prime}_{\lambda}$ to $e \delta G_{\lambda}$ is given by
\begin{equation}
\label{inelastic_ratio}
\frac{\Delta S^{\prime}_{\lambda}}{e \Delta G_{\lambda}} = \frac{8(1-\rho)\mathcal{T}^2 + (10\rho-8)\mathcal{T}+(1-2\rho)}{1-2\mathcal{T}-2(1-\mathcal{T})\rho},
\end{equation}
where $\rho=\cos^2 \theta_{\lambda}$. 
Since the out-of-phase longitudinal mode in the two-site model of Ref.~\onlinecite{PRB2012Avriller} is even mirror-symmetric, it is readily noticed that $\rho=\cos^2 \theta_{\lambda}=0$.
Thus Eq.~(\ref{inelastic_ratio}) gives the same expression as the ratio obtained from the two-site model in Ref.~\onlinecite{PRB2012Avriller},
\begin{equation}
\label{even_ratio}
\frac{\Delta S^{\prime}_{\lambda}}{e \Delta G_{\lambda}} \underset{\textrm{even}}{\longrightarrow} \frac{8\mathcal{T}^2 -8\mathcal{T}+1}{1-2\mathcal{T}}.
\end{equation}
Our theory also predicts the ratio $\Delta S^{\prime}_{\lambda}/e \Delta G_{\lambda}$ for odd vibrational modes, which is not investigated in Ref.~\onlinecite{PRB2012Avriller}:
\begin{equation}
\label{odd_ratio}
\frac{\Delta S^{\prime}_{\lambda}}{e \Delta G_{\lambda}} \underset{\textrm{odd}}{\longrightarrow} 1-2\mathcal{T}.
\end{equation}
In Fig.~\ref{fig_ratio}, purple dashed line and blue solid one represent Eqs.~(\ref{even_ratio}) and (\ref{odd_ratio}) respectively in a high transmission regime ($0.9 \leq \mathcal{T} \leq 1.0$) that the atomic chains of Ref.~\onlinecite{PRB2012Avriller} belong to.
For general cases where the mirror symmetry is broken, $\cos^2 \theta_{\lambda}$ can have a nonzero value less than unity, {\it i.e.}, $0<\cos^2 \theta_{\lambda} < 1$. 
In this case, the ratio $\Delta S^{\prime}_{\lambda}/e \Delta G_{\lambda}$ of Eq.~(\ref{inelastic_ratio}) is located between Eqs.~(\ref{even_ratio}) and (\ref{odd_ratio}) (see the green region in Fig.~\ref{fig_ratio}). 
Our theory confirms that Eq.~(\ref{even_ratio}) is indeed an upper bound to the ratio $\Delta S^{\prime}_{\lambda}/e \Delta G_{\lambda}$ as speculated in Ref.~\onlinecite{PRB2012Avriller}. Furthermore, Eq.~(\ref{odd_ratio}) is predicted to be a lower bound to the ratio $\Delta S^{\prime}_{\lambda}/e \Delta G_{\lambda}$ in a high transmission regime.

\begin{figure}[t]
\begin{center}
\includegraphics[width=1.00\columnwidth, clip=true]{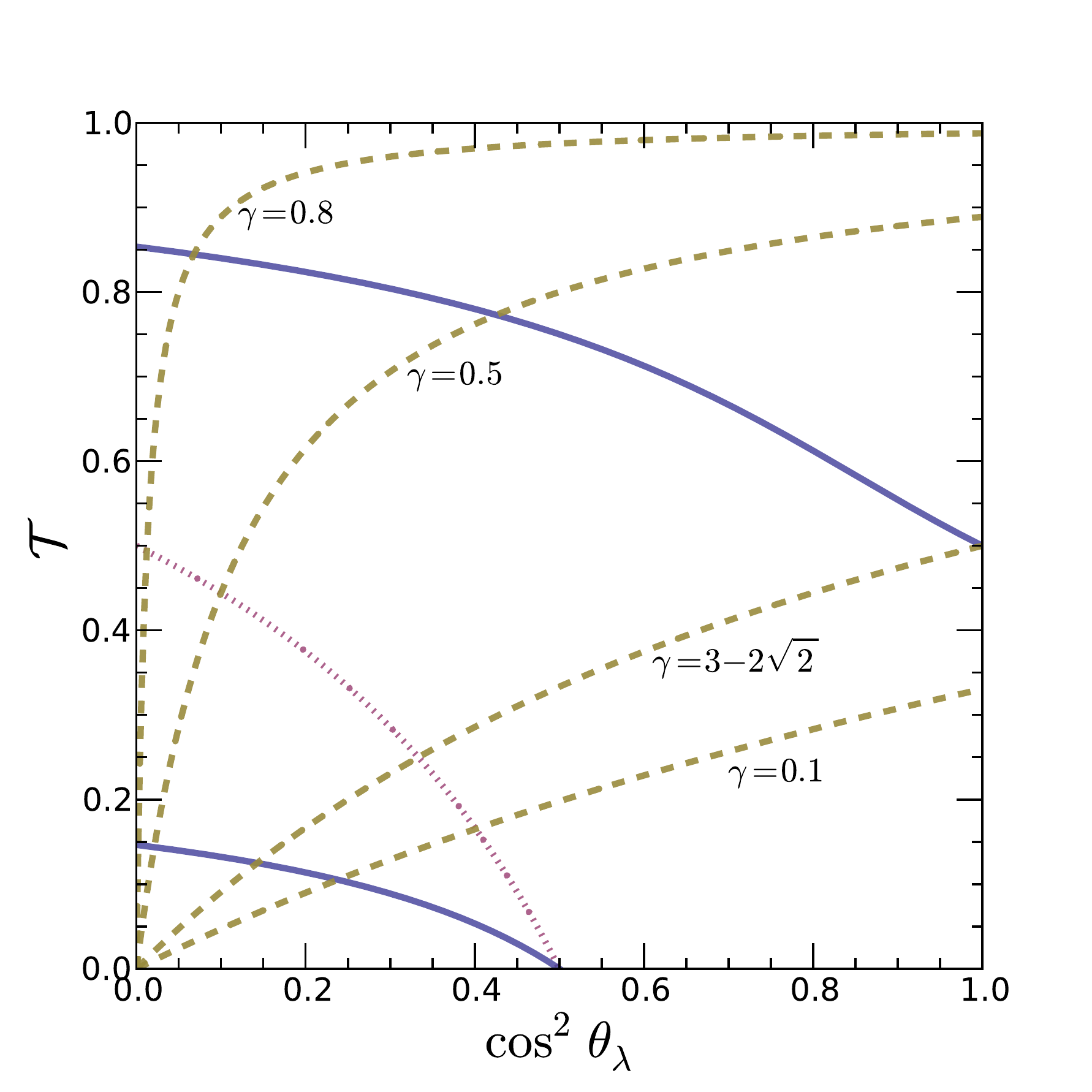} 
\end{center}
\caption{\label{fig_slm} (Color online) Schematic illustration of crossover transmissions for the non-symmetric single-level model. 
Equation~(\ref{slm_cos}) (yellow dashed lines) for $\gamma=0.1$, $3-2\sqrt{2}$, $0.5$, and $0.8$ is plotted with Eq.~(\ref{crossover}) (purple dotted line) and Eq.~(\ref{crossover_noise}) (blue solid lines). For $\gamma=3-2\sqrt{2}$, Eq.~(\ref{slm_cos}) meets the upper branch of Eq.~(\ref{crossover_noise}) at $\mathcal{T}=0.5$.}
\end{figure}

Note that the ratios $\Delta S^{\prime}_{\lambda}/e \Delta G_{\lambda}$ for some chains simulated in Ref.~\onlinecite{PRB2012Avriller} are out of the region determined by Eq.~(\ref{inelastic_ratio}).
It might be because the gold atomic chains are not strictly single-channel systems, but are multichannel with one dominant channel and a few minor channels. If it is true, our single-channel result [Eq.~(\ref{inelastic_ratio})] cannot be applied. In fact, three configurations [$\textrm{L}18.20$, $\textrm{L}19.20$, and $\textrm{L}20.20$] of Ref.~\onlinecite{PRB2012Avriller} have bare transmissions larger than unity, implying that they are multichannel. Further, some other cases such as $\textrm{L}20.50$ and $\textrm{L}26.50$ of Ref.~\onlinecite{PRB2012Avriller} have the ratios slightly outside the region determined by Eq.~(\ref{inelastic_ratio}). For these cases, our main results Eqs.~(\ref{elastic_noise1})-(\ref{elastic_noise2}), which are generalized for multichannel systems, can be used to verify possible effects of minor channels on such deviations.

\begin{figure}[b]
\begin{center}
\includegraphics[width=1.00\columnwidth, clip=true]{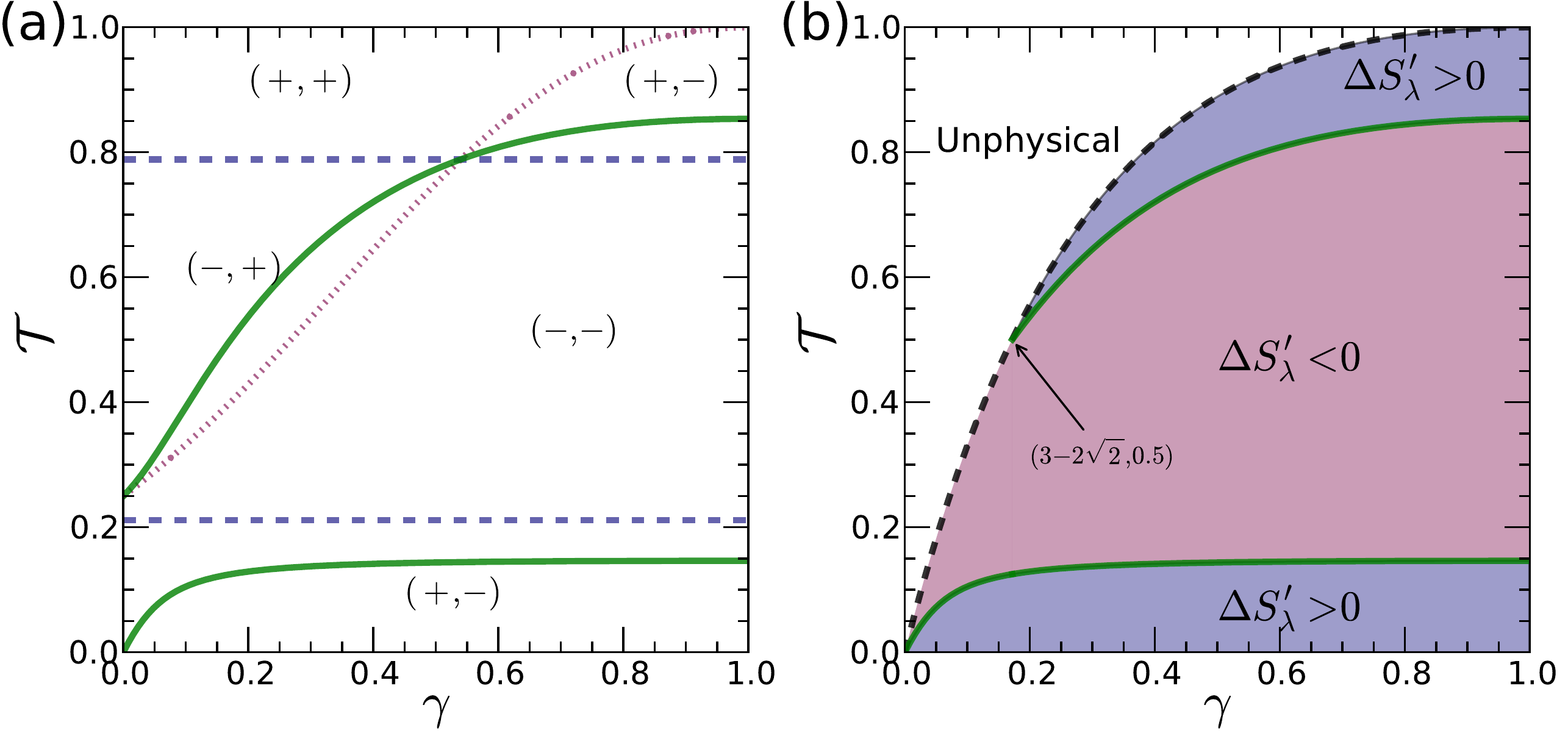} 
\end{center}
\caption{\label{slm_phase} (Color online) Phase diagram of inelastic noise signals $\Delta S^{\prime}_{\lambda}$ for the single-level model. (a) Blue dashed line and purple dotted line represent $\Delta S^{\textrm{in}\prime}_{\lambda}=0$ [Eq.~(\ref{inelastic_noise_signal})] and $\Delta S^{\textrm{el}\prime}_{\lambda}=0$ [Eq.~(\ref{elastic_noise_signal})], respectively. The first and second signs in parenthesis denote those of $\Delta S^{\textrm{in}\prime}_{\lambda}$ and $\Delta S^{\textrm{el}\prime}_{\lambda}$ respectively. Phase boundaries $\Delta S^{\prime}_{\lambda}=0$ (green solid lines) are located in the region where $\Delta S^{\textrm{in}\prime}_{\lambda}$ and $\Delta S^{\textrm{el}\prime}_{\lambda}$ have opposite signs.
(b) The phase diagram drawn together with $\mathcal{T}_{\textrm{max}}=4\gamma / (1+\gamma)^2$ (black dashed line). Blue and purple regions indicate $\Delta S^{\prime}_{\lambda}>0$ and $\Delta S^{\prime}_{\lambda}<0$, respectively. $\mathcal{T}_{\textrm{max}}$ and the upper boundary of $\Delta S^{\prime}_{\lambda}=0$ meet at $(\gamma, \mathcal{T})=(3-2\sqrt{2}, 0.5)$.}
\end{figure}
\subsection{Examples}
In the previous subsection, we discussed possible values of $\mathcal{T}^{\textrm{c}}_{\textrm{cr}}$ and $\mathcal{T}^{\textrm{c}}_{\textrm{cr}}$ without specifying $\cos^2 \theta_{\lambda}$ that is generally a function of $\mathcal{T}$, 
and we considered the mirror-symmetric cases where $\cos^2 \theta_{\lambda}$ is either $0$ or $1$. 
In this subsection, we consider general situations where the mirror symmetry is broken.
When $\cos^2 \theta_{\lambda}$ is specified for a given system, crossover transmissions $\mathcal{T}^{c}_{\textrm{cr}}$ and $\mathcal{T}^{n}_{\textrm{cr}}$ can be obtained by calculating points where 
Eqs.~(\ref{crossover}) and (\ref{crossover_noise}) intersect with $\cos^2 \theta_{\lambda}$ respectively.
We take two simple models, (1) the single-level model and (2) the $N=1$ atomic chain, 
to illustrate how $\cos^2 \theta$ depends on $\mathcal{T}$ and other parameters, and how crossover transmissions are determined.

\subsubsection{Single-level model}
First we revisit the single-level model coupled to a single local vibration~\cite{PRL2008Paulsson, PRB2009Avriller, PRB2009Schmidt, PRL2009Haupt, PRB2013SKim}. 
It is shown that $\cos^{2} \theta$ is given by 
\begin{eqnarray}
\cos^2 \theta  &=& \frac{\left(\Gamma_{R}-\Gamma_{L}\right)^2}{4\varepsilon^2+\left(\Gamma_{R}-\Gamma_{L}\right)^2} \\
\label{slm_cos}&=& \frac{\mathcal{T}}{1-\mathcal{T}}\frac{\left(1-\gamma\right)^2}{4\gamma},
\end{eqnarray}
where $\gamma \equiv \Gamma_{R}/\Gamma_{L}$ is a dimensionless parameter measuring a relative strength of couplings to left and right electrodes. Note that the maximum transmission is given by $\mathcal{T}_{\textrm{max}} = 4\gamma / (1+\gamma)^2$.
In fact, the constraint $\cos^2 \theta \leq 1$ of Eq.~(\ref{slm_cos}) leads to $\mathcal{T} \leq \mathcal{T}_{\textrm{max}}$. 

Crossing points of Eqs.~(\ref{crossover}) and (\ref{crossover_noise}) with Eq.~(\ref{slm_cos}) are crossover transmissions for the conductance and the noise signal, respectively.
Figure~\ref{fig_slm} illustrates how Eq.~(\ref{slm_cos}) intersects with Eqs.~(\ref{crossover}) and (\ref{crossover_noise}) for $\gamma$=$0.1$, $3-2\sqrt{2}$, $0.5$ and $0.8$. 
Note that Eqs.~(\ref{slm_cos}) and (\ref{crossover}) cross once, thereby implying that there is only one crossover for the conductance step. In contrast, the number of the crossovers for the inelastic noise signal can change, depending on $\gamma$. 
When $\gamma \geq 3-2\sqrt{2}$, there are two crossing points of Eqs.~(\ref{crossover_noise}) and (\ref{slm_cos}). On the other hand, when $\gamma < 3-2\sqrt{2}$, Eq.~(\ref{slm_cos}) intersects once only with the lower branch of Eq.~(\ref{crossover_noise}). 

All of these observations are well consistent with phase diagrams of the single level model~\cite{PRL2008Paulsson, PRB2009Avriller, PRB2013SKim}. Here we re-draw the phase diagram for the inelastic noise signal $\Delta S^{\prime}_{\lambda}$ in Fig.~\ref{slm_phase}, emphasizing signs of elastic and inelastic contributions $\Delta S^{\textrm{el}\prime}_{\lambda}$ and $\Delta S^{\textrm{in}\prime}_{\lambda}$. 
For $\gamma>3-2\sqrt{2}$, it is shown that there are two phase boundaries (green solid lines) separating $\Delta S^{\prime}_{\lambda}>0$ and $\Delta S^{\prime}_{\lambda}<0$ in Fig.~\ref{slm_phase}(b). The upper phase boundary of $\Delta S^{\prime}_{\lambda}=0$ ends at $(\gamma, \mathcal{T})=(3-2\sqrt{2}, 0.5)$ where it meets $\mathcal{T}_{\textrm{max}}$. For $\gamma<3-2\sqrt{2}$, there is only one phase boundary in a low transmission regime. 



\subsubsection{$N$=$1$ atomic chain}
\begin{figure}[b]
\begin{center}
\includegraphics[width=0.9\columnwidth, clip=true]{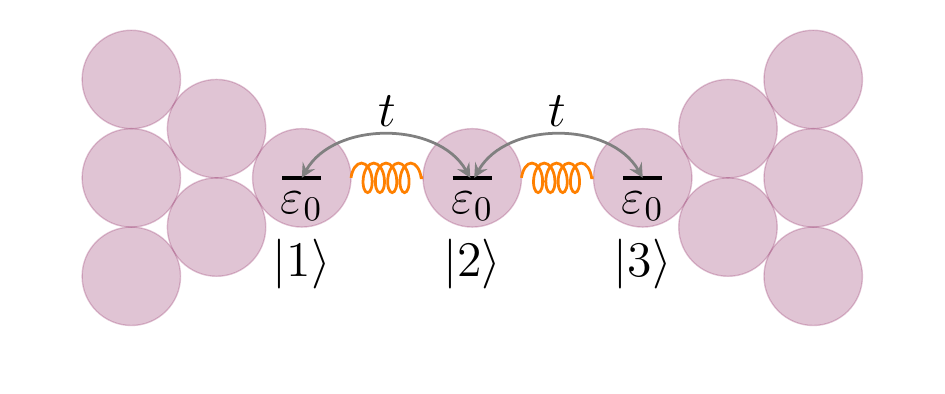} 
\end{center}
\caption{\label{conf_atc} (Color online) Schematic illustration of the $N$=$1$ atomic chain model. Three orbitals $|1\rangle$, $|2\rangle$, and $|3\rangle$ constitute the device region. $t$ is a hopping parameter bewteen nearest neighbors. The central atom is allowed to vibrate.}
\end{figure}


\begin{figure}[t]
\begin{center}
\includegraphics[width=1.0\columnwidth, clip=true]{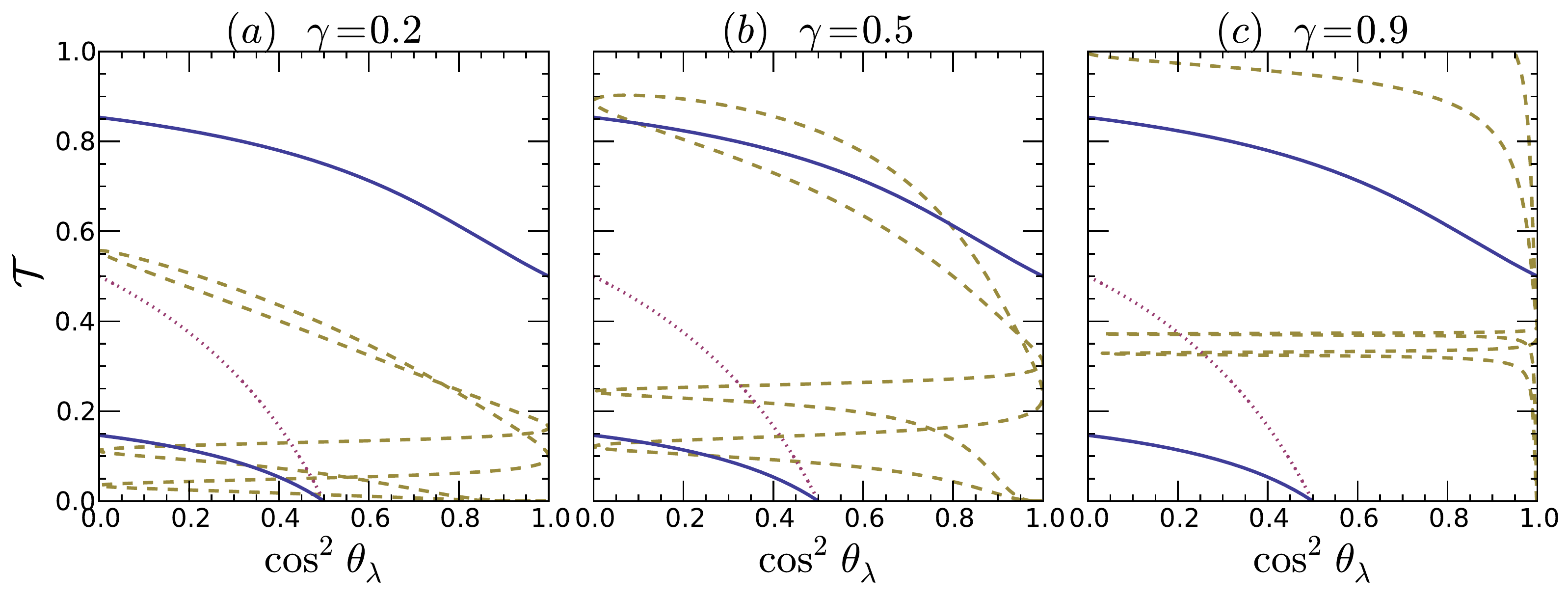} 
\end{center}
\caption{\label{fig_atc} (Color online) Crossover transmissions of the conductance and the shot noise for the $N$=$1$ chain model: (a) $\gamma=0.2$, (b) $\gamma=0.5$, and (c) $\gamma=0.9$. Purple dotted line, blue solid line and yellow dashed line denote trajectories of Eqs.~(\ref{crossover}), (\ref{crossover_noise}), and (\ref{atc_cos}), respectively. }
\end{figure}

As the second example, we consider the tight-binding model for the shortest atomic chain, {\it i.e.}, the $N$=$1$ chain, 
which can be a simple model with odd mirror-symmetric vibrations. The central region consists of three orbitals: 
$|1\rangle$ and $|3\rangle$ are located at outermost atoms of electrodes and $|2\rangle$ is at a central atom vibrating between electrodes. 
The corresponding Hamiltonian is
\begin{equation}
H_{C} = \sum_{i=1}^{3} \varepsilon_{0} d_{i}^{\dagger}d_{i} + \sum_{i=1,2} \left[ t d_{i}^{\dagger}d_{i+1} + \textrm{H.c.} \right],
\end{equation}
where $d_{i}$ and $d_{i}^{\dagger}$ are electronic annihilation and creation operators of the $i$th site, respectively. 
The hopping parameter $t$ is taken to be real for simplicity. 
Considering a longitudinal motion of the central atom, its vibrational mode is odd mirror-symmetric. 
The el-vib coupling Hamiltonian is written as
\begin{equation}
\mathcal{H}_{el-vib} = \mathcal{M} \sum_{i=1,2}\left[ (-1)^{i+1} d_{i}^{\dagger} d_{i+1} + \textrm{H.c.} \right] \left(a+a^{\dagger}\right)
\end{equation}
When left and right coupling functions are simply given by $\Gamma_{L}|1\rangle \langle1|$ and $\Gamma_{R}|3\rangle \langle3|$ respectively, 
one can show that
\begin{eqnarray}
\label{atc_cos}\cos^2 \theta_{\lambda} &=& \frac{\left( \Delta \tilde{\varepsilon}^3 - 2\Delta \tilde{\varepsilon} \tilde{t}^2 + \frac{1}{4} \Delta \tilde{\varepsilon} \gamma \right)^2}{\left(\Delta \tilde{\varepsilon}^3 -2\Delta \tilde{\varepsilon} \tilde{t}^2 + \frac{1}{4}\Delta \tilde{\varepsilon} \gamma\right)^2 + \frac{1}{4}\left(\Delta \tilde{\varepsilon}^2 - \tilde{t}^2 \right)^2 \left(1-\gamma\right)^2}, \nonumber \\
\end{eqnarray}
and  
\begin{eqnarray}
\label{atc_T}\mathcal{T}   &=& \frac{\gamma \tilde{t}^4}{\left|\left(\Delta \tilde{\varepsilon} + \frac{i}{2}\right)^2\left(\Delta \tilde{\varepsilon} + \frac{i}{2}\gamma\right)\Delta \tilde{\varepsilon} - \tilde{t}^2\left(2\Delta \tilde{\varepsilon} + \frac{i}{2}[1+\gamma] \right)\right|^2}, \nonumber \\ 
\end{eqnarray} 
where $\Delta \tilde{\varepsilon} = (\varepsilon_{F} - \varepsilon_{0})/\Gamma_{L}$, $\tilde{t} = t/\Gamma_{L}$, and $\gamma=\Gamma_{R}/\Gamma_{L}$.
$\cos^2 \theta_{\lambda}$ [Eq.~(\ref{atc_cos})] implicitly depends on $\mathcal{T}$ [Eq.~(\ref{atc_T})] via system parameters such as $\Delta \tilde{\varepsilon}$, $\tilde{t}$, and $\gamma$. 

For the symmetric coupling $\gamma=1$, Eq.~(\ref{atc_cos}) becomes unity ($\cos^2 \theta_{\lambda}$=$1$) as predicted in the previous subsection. Note that $\cos^2 \theta_{\lambda}$ is indeterminate at $\Delta \tilde{\varepsilon}=0, \pm\sqrt{2\tilde{t}^2-0.25}$. Since $r^\prime t \mathbb{M}^{\lambda}_{R L} =0$ for $\Delta \tilde{\varepsilon}=0, \pm\sqrt{2\tilde{t}^2-0.25}$, $\theta_{\lambda}$ is not defined. 
In this perfect transmission case, one can directly calculate $\Delta G_{\lambda}$ and $\Delta S_{\lambda}^{\prime}$
by using original expressions, Eqs.~(\ref{elasticT1}), (\ref{elasticT2}), and (\ref{inelastic_noise1})-(\ref{elastic_noise2}). 

Next we consider non-symmetric cases, {\it i.e.}, $\gamma \neq 1$.
For simplicity, we assume that non-symmetric electrode couplings do not alter the hopping parameter $t$ and the vibrational mode in the conductor region.
In Fig.~\ref{fig_atc}, trajactories of Eqs.~(\ref{atc_cos}) and (\ref{atc_T}) (yellow dashed lines) are plotted by tuning the parameter $\Delta \tilde{\varepsilon}$ for $\gamma=0.2$, $0.5$, and $0.9$.
Here $\tilde{t}$ is $0.5$. Figure~\ref{fig_atc} illustrates how crossover transmissions for noise signals qualitatively change as $\gamma$ varies. 
For $\gamma=0.2$, $\cos^2 \theta_{\lambda}$ intersects four times only with the lower branch of Eq.~(\ref{crossover_noise}).
When we increase $\gamma$ to $0.5$, $\cos^2 \theta_{\lambda}$ meets both upper and lower branches of Eq.~(\ref{crossover_noise}).
As $\gamma$ approaches $1$, crossing points with the lower branch disappear, and crossover transmissions of the upper branch become closer to $0.5$, which is the crossover transmission 
for the odd mirror-symmetric case. 

We also notice that $\Delta \tilde{\varepsilon}=0$, which implies the particle-hole symmetry, is of particular interest for non-symmetric junctions ($\gamma \neq 1$).
In this case, $\cos^2 \theta_{\lambda}$ always vanishes, irrespective of $\gamma(<1)$ and $\tilde{t}$, and thus it is expected that $\mathcal{T}_{cr}^{c}=0.5$ and $\mathcal{T}_{cr}^{n}=(2\pm\sqrt{2})/4$ for any non-symmetric coupling. 

\section{Conclusions}\label{conclusion}
In this paper, we have presented the scattering-state description of the inelastic shot noise in a regime of a weak el-vib coupling and equilibrated vibrons. As discussed in the inelastic current, the inelastic shot noise is determined by the interplay of elastic and inelastic scattering processes. The elastic and inelastic contributions to the current noise are further decomposed into current correlations of electrons at the same energy and those of electrons at two energies that differ by the vibrational energy $\hbar \omega_{\lambda}$. 

Applied to single-channel systems, our description enables to find out two ranges of transmission at which the crossover in the inelastic noise signal can take place. In particular, for mirror-symmetric systems, we have shown that even parity modes lead to the crossover at $\mathcal{T}=(2\pm\sqrt{2})/4$, while the crossover occurs at $\mathcal{T}=0.5$ for odd parity ones. 
Considering the ratio $\Delta S^{\prime}_{\lambda}/e \delta G_{\lambda}$, we have confirmed that the ratio $\Delta S^{\prime}_{\lambda}/e \delta G_{\lambda}$ of the even parity mode is indeed an upper bound to ratios of general cases in a high transmission regime as speculated in Ref.~\onlinecite{PRB2012Avriller}, and further we have predicted that the ratio $\Delta S^{\prime}_{\lambda}/e \delta G_{\lambda}$ of the odd parity mode is a lower bound. 

Our scattering-state description is formulated for general situations involving many electronic states, many vibrational modes, and multiple conducting channels, so that it can be used to analyze first-principle calculation results, especially when specification of inter-channel and intra-channel scattering processes is crucial to understand the results.  


\section{acknowledgement}
We thank Prof. Mahn-Soo Choi for helpful discussions. This work is supported by the CAC of KIAS.

\appendix
\section{Derivation of Eqs.~(\ref{inelastic_noise1})-(\ref{elastic_noise2})}\label{derivation}
Using Eqs.~(\ref{identity1})-(\ref{identity2-2}), one can show that 
\begin{widetext}
\begin{eqnarray}
\frac{\delta S_{\textrm{mf}}}{2e^2/h} &=& (2\pi)^2 \sum_{\lambda, m,n}\int^{\mu_{L}}_{\mu_{R+\hbar\omega_{\lambda}}} d\varepsilon \left[(1-2\mathcal{T}_{m})\mathcal{R}_{m} - (1-2\mathcal{T}^{-}_{n})\mathcal{T}^{-}_{n} \right] \left|\langle \Phi_{Rn}^{-}|\mathcal{M}^{\lambda}|\Phi_{Lm}\rangle \right|^2 \\
&&-\left(2\pi\right)^{2}\sum_{\lambda,m,n}\int_{\mu_{R}+\hbar\omega_{\lambda}}^{\mu_{L}}d\varepsilon\left(1-2\mathcal{T}_{m}\right)\textrm{Re}\left[r_{m}^{\prime}t_{m}^{*}\langle\Phi_{Rm}|\mathcal{M}^{\lambda}|\Phi_{Rn}^{-}\rangle\langle\Phi_{Rn}^{-}|\mathcal{M}^{\lambda}|\Phi_{Lm}\rangle\right] \\
&&+\left(2\pi\right)^{2}\sum_{\lambda,m,n}\int_{\mu_{R}}^{\mu_{L}-\hbar\omega_{\lambda}}d\varepsilon\left(1-2\mathcal{T}_{m}\right)\textrm{Re}\left[r_{m}^{\prime}t_{m}^{*}\langle\Phi_{Rm}|\mathcal{M}^{\lambda}|\Phi_{Ln}^{+}\rangle\langle\Phi_{Ln}^{+}|\mathcal{M}^{\lambda}|\Phi_{Lm}\rangle\right]
\end{eqnarray}
and
\begin{eqnarray}
\frac{\delta S_{\textrm{vc}}}{2e^2/h} &=& (2\pi)^2 \sum_{\lambda, m,n}\int^{\mu_{L}}_{\mu_{R+\hbar\omega_{\lambda}}} d\varepsilon 2(\mathcal{T}_{m}-1) \mathcal{T}^{-}_{n} \left|\langle \Phi_{Rn}^{-}|\mathcal{M}^{\lambda}|\Phi_{Lm}\rangle \right|^2 \\
&&+2\left(2\pi\right)^{2}\sum_{\lambda,m,n}\int_{\mu_{R}+\hbar\omega_{\lambda}}^{\mu_{L}}d\varepsilon\textrm{Re}\left\{ r_{m}^{\prime}t_{m}^{*}r_{n}^{\prime-}t_{n}^{-*}\langle\Phi_{Rm}|\mathcal{M}^{\lambda}|\Phi_{Ln}^{-}\rangle\langle\Phi_{Rn}^{-}|\mathcal{M}^{\lambda}|\Phi_{Lm}\rangle\right\}.
\end{eqnarray}
The prefactors of transition amplitudes $\left|\langle \Phi_{Rn}^{-}|\mathcal{M}^{\lambda}|\Phi_{Lm}\rangle \right|^2$ in $\delta S_{\textrm{mf}}$ and $\delta S_{\textrm{mf}}$ can be written as follows:
\begin{equation}
(1-2\mathcal{T}_{m})\mathcal{R}_{m} - (1-2\mathcal{T}^{-}_{n})\mathcal{T}^{-}_{n} = (1-2\mathcal{T}_{n}^{-})\mathcal{R}^{-}_{n} - (1-2\mathcal{T}_{m})\mathcal{T}_{m} + 2(\mathcal{T}_{n}^{-} - \mathcal{T}_{m})
\end{equation}
and
\begin{equation}
2(\mathcal{T}_{m}-1) \mathcal{T}^{-}_{n} = -2\mathcal{R}_{n}^{-} \mathcal{T}_{m} - 2(\mathcal{T}_{n}^{-} - \mathcal{T}_{m})
\end{equation}
Thus, by canceling out terms proportional to $2(\mathcal{T}_{n}^{-} - \mathcal{T}_{m})$ in $\delta S_{\textrm{mf}}$ and $\delta S_{\textrm{mf}}$, one can finally obtain Eqs.~(\ref{inelastic_noise1})-(\ref{elastic_noise2}). 

\section{Multichannel Systems}\label{multichannel_systems}
For multi-channel systems, corrections to the conductance and the noise are decomposed into {\it intra}-channel and {\it inter}-channel scattering contributions,
\begin{eqnarray}
\delta G &=& \delta G_{\textrm{intra}} + \delta G_{\textrm{inter}} \\
\Delta S^\prime &=& \Delta S^\prime_{\textrm{intra}} + \Delta S^\prime_{\textrm{inter}}.
\end{eqnarray}
The intra-channel scattering contributions are simply a collection of each eigenchannel contribution, which is the same form of the single-channel result:
\begin{eqnarray}
\frac{\delta G_{\textrm{intra}}}{2e^2/h} &=& (2\pi)^2 \sum_{\lambda,m} \left(1 -2\mathcal{T}_{m}-2\mathcal{R}_{m} \cos^2 \theta_{\lambda m} \right) \left|\mathbb{M}^{\lambda m}_{RL} \right|^2 \\
\frac{\Delta S^\prime_{\textrm{intra}}}{2e^3/h} &=& (2\pi)^2 \sum_{\lambda, m} \left[\left(8-8\cos^2 \theta_{\lambda m}\right)\mathcal{T}^2 + \left(10\cos^2 \theta_{\lambda m} - 8\right)\mathcal{T} + \left(1-2\cos^2 \theta_{\lambda m}\right)\right]\left|\mathbb{M}^{\lambda m}_{RL} \right|^2,
\end{eqnarray}
where $\mathbb{M}^{\lambda m}_{RL} = \langle \Phi_{Rm} | \mathcal{M}^{\lambda} |\Phi_{Lm}\rangle$ and $\theta_{\lambda m} = \arg \left[r^\prime_{m} t^{*}_{m} \mathbb{M}^{\lambda m}_{RL} \right]$.
Unfortunately, the inter-channel scattering terms are not simplified as the intra-channel ones. 
When the mirror symmetry along the transport direction is respected, the inter-channel scattering contributions can be written in the form of Fermi's golden rule:
\begin{eqnarray}
\frac{\delta G_{\textrm{inter}}}{2e^2/h} &=& \left(2\pi\right)^{2}\sum_{\lambda,m\neq n}\frac{2\sqrt{\mathcal{R}_{m}\mathcal{T}_{m}}}{\sqrt{\mathcal{R}_{n}\mathcal{T}_{m}}\pm\sqrt{\mathcal{R}_{m}\mathcal{T}_{n}}e^{-i\left(\Delta\theta_{m}+\Delta\theta_{n}\right)}} \cos\left(\theta_{m}^{r}-\theta_{n}^{r}+2\varphi_{mn}^{\lambda}\right) \left|\langle R_{m}|\mathcal{M}^{\lambda}|L_{n}\rangle\right|^{2} \\
\frac{\Delta S^{\prime}_{\textrm{inter}}}{2e^2/h} &=& (2\pi)^2\sum_{\lambda,m\neq n}2\left(1-2\mathcal{T}_{m}\right)\frac{\sqrt{\mathcal{R}_{m}\mathcal{T}_{m}}}{\sqrt{\mathcal{R}_{n}\mathcal{T}_{m}}\pm\sqrt{\mathcal{R}_{m}\mathcal{T}_{n}}e^{-i\left(\Delta\theta_{m}+\Delta\theta_{n}\right)}}\cos\left(\theta_{m}^{r}-\theta_{n}^{r}+2\varphi_{mn}^{\lambda}\right)\left|\langle R_{m}|\mathcal{M}^{\lambda}|L_{n}\rangle\right|^{2} \nonumber\\
&&+(2\pi)^2\sum_{\lambda,m \neq n}\left(1-2\mathcal{T}_{m}\right)\frac{-\sqrt{\mathcal{R}_{n}\mathcal{T}_{m}}\pm\sqrt{\mathcal{R}_{m}\mathcal{T}_{n}}e^{-i\left(\Delta\theta_{m}+\Delta\theta_{n}\right)}}{\sqrt{\mathcal{R}_{n}\mathcal{T}_{m}}\pm\sqrt{\mathcal{R}_{m}\mathcal{T}_{n}}e^{-i\left(\Delta\theta_{m}+\Delta\theta_{n}\right)}}\left|\langle R_{m}|\mathcal{M}^{\lambda}|L_{n}\rangle\right|^{2} \nonumber\\
&&-2(2\pi)^2\sum_{\lambda, m \neq n}\mathcal{R}_{n}\mathcal{T}_{m}\left|\langle R_{m}|\mathcal{M}^{\lambda}|L_{n}\rangle\right|^{2} \pm 2(2\pi)^2\sum_{\lambda,m \neq n}\sqrt{\mathcal{R}_{m}\mathcal{T}_{m}\mathcal{R}_{n}\mathcal{T}_{n}}e^{i\left(\Delta\theta_{m}+\Delta\theta_{n}\right)}\left|\langle R_{m}|\mathcal{M}^{\lambda}|L_{n}\rangle\right|^{2},
\end{eqnarray}
where the upper (lower) sign is for even (odd) mirror-symmetric modes. Here $\theta^{r}_{m} = \arg \left[r_{m} \right]$, $\theta^{t}_{m} = \arg \left[t_{m} \right]$, $\Delta \theta_{m} = \theta^{r}_{m} - \theta^{t}_{m}$, and $\varphi_{mn}^{\lambda}=\arg\left[\langle R_{m}|\mathcal{M}|L_{n}\rangle\right]$.
Note that $r_{m}t_{m}^{*}+r_{m}^{*}t_{m}=0$ for the unitarity of the scattering matrix. It implies that $r_{m} t_{m}^{*}$ is purely imginary, and $\exp\left[i\Delta \theta_{m} \right]=\pm 1$.
\end{widetext}

\end{document}